\documentclass[a4paper,11pt]{article}
\pdfoutput=1 
\usepackage{jheppub} 
\usepackage{amsmath,amssymb,epsfig,relsize,xspace,slashed,braket,xcolor,mathtools}

\allowdisplaybreaks
 
\definecolor{nicered}{rgb}{0.7,0.1,0.1}
\definecolor{nicegreen}{rgb}{0.1,0.5,0.1}
\definecolor{niceblue}{rgb}{0.0,0.1,0.7}
\hypersetup{colorlinks,citecolor=nicegreen,linkcolor=nicered,urlcolor=niceblue}

\newcommand{\overbar}[1]{\mkern 3.5mu\overline{\mkern-3.5mu#1\mkern-1.mu}\mkern 3.5mu}
\DeclareMathOperator{\Tr}{Tr}
\DeclareMathOperator{\im}{Im}

\title{Light-cone sum rules for proton decay}

\author[1]{Ulrich Haisch}

\author[1]{and Amando Hala}

\affiliation[1]{Max Planck Institute for Physics, F{\"o}hringer Ring 6,  80805 M{\"u}nchen, Germany}   

\emailAdd{haisch@mpp.mpg.de}
\emailAdd{ahala@mpp.mpg.de}

\abstract{We estimate the form factors that parametrise the hadronic matrix elements of proton-to-pion transitions with the help of light-cone sum rules. These form factors are relevant for  semi-leptonic proton decay channels induced by baryon-number violating dimension-six operators,  as typically studied in the context of grand unified theories. We calculate the form factors in a kinematical regime where the momentum transfer from the proton to the pion is space-like and extrapolate our final results to the regime that is relevant for proton decay. In this way, we obtain estimates for the form factors that show  agreement with the state-of-the-art calculations in lattice QCD, if systematic uncertainties are taken into account. Our work is a first step towards calculating more involved proton decay channels where lattice~QCD results are not available at present.}

\begin{document} 

\maketitle

\section{Introduction}
\label{sec:intro}

Proton decay would be a clear indication of physics beyond the Standard Model~(SM) if it was measured. While the global symmetries of the~SM forbid proton decay, these symmetries arise accidentally. In fact, considering baryon-number violation at the perturbative level is well motivated by theories of grand unification~(GUTs)~\cite{Georgi:1974sy,Fritzsch:1974nn}, supersymmetric theories~\cite{Nilles:1983ge,Haber:1984rc,Chamoun:2020aft}, models of baryogenesis~\cite{Sakharov:1967dj,Canetti:2012zc} and more generally in theories of quantum gravity,  where the global symmetries of the~SM are expected to be broken at some level~\cite{Banks:2010zn,Harlow:2018jwu}.  In~the case of GUTs, baryon number is typically violated by tree-level interactions, and proton decay is mediated by the massive gauge fields of the spontaneously broken unified gauge group~\cite{Langacker:1980js,Raby:2002wc,Nath:2006ut,Ellis:2019fwf}. On the experimental side,  searches for simple proton decay channels, such as decays into a pseudoscalar meson and an anti-lepton, provide very strong constraints on the proton lifetime of $\tau_p \gtrsim 10^{34}$ years~\cite{Miura:2016krn}. These bounds allow one to probe theories that predict proton decay up to extremely high energy scales,  thereby  putting severe constraints on the scale of unification.

Obtaining such constraints requires a theoretical prediction for the  widths of  semi-leptonic proton decay channels, which in turn relies on the knowledge of the hadronic matrix element of the underlying proton-to-meson transition. Water-Cherenkov experiments such as Super-Kamiokande~\cite{Abe:2013gga,Fukuda:2002uc} attempt to measure the decay products of protons that are approximately at rest, and thus the relevant energy scale for the hadronic transition is given by the proton mass. A perturbative description of the relevant  hadronic matrix elements in QCD is not possible at this energy scale because of large radiative corrections due to the exchange of soft gluons. A prediction for the hadronic matrix elements by other means is  therefore required to probe baryon-number violating new physics with the help of experimental data from proton decay searches.

Early attempts to compute the hadronic matrix elements   of semi-leptonic proton decays date back to the '80s and employed non-relativistic quark models, often based on the approximate   $SU(6)$ flavour-spin symmetry  of the partons~\cite{Jarlskog:1978uu,Machacek:1979tx,Goldman:1980ah,Gavela:1981cf,Salati:1982bw}, bag models which allow for relativistic partons~\cite{Din:1979bz,Donoghue:1979pr,Golowich:1980ne,Donoghue:1982jm,Wakano:1982sk,Okazaki:1982eh}, or QCD sum rules~\cite{Berezinsky:1981qb}. Also an effective chiral theory was proposed in  the articles~\cite{Claudson:1981gh,Isgur:1982iz,Chadha:1983sj,Kaymakcalan:1983uc}, which can be used to derive relations among the various two-body decay widths but still contains a priori unknown low-energy constants.  As a result  the latter approach cannot predict the absolute value of the proton decay width without further input. The methods mentioned above have also been applied to  estimate these low-energy constants, in which case the final predictions for the  hadronic matrix elements suffer from additional systematic uncertainties due the approximate nature of the effective chiral theory. Moreover, the results of these  model calculations differ by up to an order of magnitude from each other (see Table~VI in~\cite{Aoki:2006ib} for a summary and comparison). On the other hand, lattice~QCD~(LQCD) groups have by now achieved to directly compute the   needed hadronic matrix elements within uncertainties of $(10-15)\%$~\cite{Gavela:1988cp,Aoki:1999tw,Tsutsui:2004qc,Aoki:2006ib,Braun:2008ur,Aoki:2013yxa,Aoki:2017puj,Yoo:2018fyn}. These results cover all two-body decays into pseudoscalar mesons and light anti-leptons, which are relevant for GUTs. 

Experimental bounds are today available for a broad range of baryon-number violating processes~\cite{Takhistov:2016eqm,Heeck:2019kgr,Girmohanta:2019xya}. In particular, inclusive proton {($p$) decay searches  for processes like  $p \rightarrow \pi^0\, \ell^+ + X$ that involve a neutral pion~($\pi^0$) and an anti-lepton~($\ell^+$) might be of interest if baryon-number violation does not become manifest in a simple two-body decay. For~example, the case where $X$ is a graviton  may provide relevant constraints on  theories where baryon-number violation occurs in connection with gravity such as in the effective theory of gravitons and~SM particles called GRSMEFT~\cite{Ruhdorfer:2019qmk,Durieux:2019siw}.  LQCD results are not available for processes of this kind, which raises the question: how can one obtain estimates of the proton  lifetime in such cases?

In this work, we  establish  a method that allows to estimate the hadronic matrix elements that enter processes of the type $p \rightarrow \pi^0\, \ell^+ \, (+ \, X)$. As a proof-of-principle we apply our general approach to the simple  two-body  case $p \rightarrow \pi^0\, e^+$ with $e^+$ a positron, leaving  the application to three-body proton decay processes such as transitions involving an additional graviton for future work. In fact, studying the simple decay mode $p \rightarrow \pi^0\, e^+$ allows us to make a thorough comparison with the latest LQCD results~\cite{Aoki:2017puj}. In this way we are not only able to validate our method but can also assess the systematic uncertainties that plague our estimates.  Our method employs the techniques of light-cone sum rules~(LCSRs) in QCD. In~particular, we perform an operator product expansion~(OPE) on the light-cone, which allows us to factorise the hard scattering from the soft interactions. While the hard-scattering kernel can be computed perturbatively in QCD, the soft contributions are parametrised in terms of condensates and distribution amplitudes~(DAs) of the final-state pion, which  enter the LCSRs  as input. The light-cone expansion works well if the momentum transfer $q$ from the proton to the pion is large in magnitude and space-like,~i.e.~$q^2 < 0$.  We therefore cannot directly compute the hadronic matrix elements at the physical point of the two-body decay kinematics, where $q^2$ is fixed and equal to the square of the positron mass. However, we are able to find values in the space-like regime at $q^2 \simeq -0.5 \, \rm{GeV}^2$, which are close enough to the physical regime to provide an estimate of the hadronic matrix elements at the physical point by means of suitable extrapolations.  Albeit~our approach does not  achieve the same level of accuracy  as the state-of-the-art LQCD calculation~\cite{Aoki:2017puj}, we believe that the results of this work are promising, because the obtained precision is better  than the methods that have been developed in the '80s to estimate proton decay rates. Furthermore, the LCSR approach developed by us in this article should be able to  at least provide order-of-magnitude estimates for hadronic matrix elements that enter certain three-body proton decay processes. Such decays could be phenomenologically relevant (see for instance~\cite{Heeck:2019kgr}) but only model estimations exist for selected modes~\cite{Wise:1980ch}, making three-body final-state proton decay processes  an interesting target of future LQCD studies~\cite{Cirigliano:2019jig}. 

 Our work is organised is as follows. In Section~\ref{sec:pheno} we put the discussion of hadronic matrix elements for the $p \rightarrow \pi^0\,  e^+$ decay on a more systematic footing. In particular, we use all of the dimension-six operators in the so-called~SM effective field theory~(SMEFT) that are relevant for this decay, so that our analysis is model-independent. These operators are typically generated by baryon-number violating new physics that can be integrated out below a certain (large) energy scale. After that we decompose the hadronic matrix elements into form factors. These form factors enter a correlation function that is computed with the help of LCSR techniques in Section~\ref{sec:lcsr},  which allows us to derive the LCSRs for the form factors relevant for proton decay in GUTs. In Section~\ref{sec:analysis} we turn to the numerical evaluation of the  LCSRs  and compute the form factors in the regime of virtual momentum transfer. Eventually, we compare our findings to the results of the latest LQCD computation~\cite{Aoki:2017puj} and discuss in detail the uncertainties that enter our final estimates. We~conclude and present an outlook in Section~\ref{sec:conclusions}.  Technical details are relegated to three appendices.

\section{Phenomenological parametrisation}
\label{sec:pheno}

Integrating out heavy new physics that violates baryon number typically generates the following   dimension-six SMEFT operators,
\begin{equation}
\mathcal{L}^{(6)}_{\slashed{B}} = \sum_{\Gamma,\Gamma^\prime} c_{\Gamma \Gamma'} \, \mathcal{O}_{\Gamma \Gamma'} = \sum_{\Gamma,\Gamma^\prime} c_{\Gamma \Gamma'} \, \epsilon^{abc} \left(d_a^T C P_{\, \Gamma} u_b \right) \left(e^T C P_{\, \Gamma^\prime} u_c \right) \,,
\label{eq:4foperators}
\end{equation}
where $C$ is the charge conjugation matrix, $T$ denotes the transpose of the Dirac index, the symbols $P_{\, \Gamma,\Gamma^\prime}$ denote the left- and right-chiral projectors $P_L$ and $P_R$ such that $\Gamma$ ($\Gamma^\prime$) denotes the chirality of the first (second) fermion bilinear, $\epsilon^{abc}$ is the fully antisymmetric Levi-Civita tensor and $a,b,c$ are colour indices. In the following analysis we restrict ourselves to the first generation of up- and down-type quarks $u$ and $d$ and the electron~$e$. We consider all possible chirality combinations of the interaction~\eqref{eq:4foperators} in order to provide a  model-independent analysis.  Note that the Wilson coefficients $c_{\Gamma \Gamma'}$, which encode   short-distance physics, have a mass dimension of  $-2$.

The transition matrix element of the proton decay $p \rightarrow \pi^0\, e^+$ induced by an insertion of  an operator entering~(\ref{eq:4foperators}) can be factorised into a hadronic and leptonic part (up to electroweak corrections),
\begin{equation}
\braket{\pi^0(p_\pi) e^+(q) |  \mathcal{O}_{\Gamma \Gamma'}  | p(p_p)} = \bar{v}_e^c (q)\, H_{\Gamma \Gamma^\prime}( p_p, q ) u_p(p_p)\,.
\end{equation}
Here $u_p(p_p)$ denotes the spinor of the proton with momentum $p_p$ and  $\bar{v}_e^c (q)$ is the charge conjugate anti-spinor of the electron with momentum $q \equiv  p_p - p_\pi$.} The main goal of the following analysis is to calculate the hadronic matrix element  $H_{\Gamma \Gamma^\prime} ( p_p, q )$ of the $p \rightarrow \pi^0$ transition, 
\begin{equation}
H_{\Gamma \Gamma^\prime}(p_p,q)  u_p(p_p) \equiv \braket{\pi^0(p_\pi) |  \, \epsilon^{abc} \left(d_a^T C P_{\, \Gamma} u_b \right) P_{\, \Gamma^\prime} u_c \,  | p(p_p)} \,,
\end{equation}
where all the quark fields are evaluated at the  space-time point $x=0$.   For an on-shell proton the  above matrix element can be decomposed into  two form factors as follows,
\begin{equation} \label{eq:W0W1}
H_{\Gamma \Gamma^\prime}(p_p,q) u_p (p_p) = i P_{\, \Gamma^\prime} \left(W^0_{\Gamma \Gamma^\prime} (q^2) + \frac{\slashed{q}}{m_p} \, W^1_{\Gamma \Gamma^\prime}  (q^2) \right)u_p (p_p) \,,
\end{equation}
with $m_p = 938 \, {\rm MeV}$ the proton mass. Notice that the form factors are related due to parity, which is conserved in QCD. Specifically, one has
\begin{equation}
W^n_{RR} (q^2)= W^n_{LL}(q^2) \,, \qquad W^n_{LR} (q^2)= W^n_{RL}(q^2) \,,
\end{equation}
with $n=0,1$. In this work we calculate the combinations $\Gamma \Gamma^\prime = RR, LR$ explicitly, which covers all chirality combinations due to the above relations.  

The starting point for evaluating the form factors $W^n_{\Gamma \Gamma^\prime} (q^2)$ in LCSRs is the correlation function
\begin{equation}
\Pi_{\Gamma\Gamma^\prime} (p_p,q) = i \int \! d^4x \, e^{iqx} \braket{\pi^0(p_\pi)| \, T \left[ Q_{\Gamma \Gamma^\prime}(x) \bar{\eta}(0) \right] |0} \,,
\label{eq:corrfunc}
\end{equation}
where $T$ denotes time ordering and the current $\eta$ ($\bar{\eta} \equiv \eta^\dagger \gamma^0$) is a combination of three quark fields  that interpolates the proton,
\begin{equation}
\braket{0| \eta(0) | p(p_p)} =m_p \lambda_p u_p( p_p) \,.
\label{eq:protoncoupl}
\end{equation}
Here $\lambda_p$ denotes the couplings strength of the current $\eta$ to the physical proton state. The strongly-interacting parts of the  dimension-six operators~\eqref{eq:4foperators} are represented  by
\begin{equation}
Q_{\Gamma\Gamma^\prime}(x) \equiv \epsilon^{abc} \left(d_a^T(x) C P_{\, \Gamma} u_b(x) \right) P_{\, \Gamma^\prime} u_c(x) \,.
\end{equation}

To obtain a parametrisation of the hadronic matrix elements  $H_{\Gamma \Gamma^\prime} ( p_p, q )$ we insert a complete set of intermediate states that have the same quantum numbers as the proton into~(\ref{eq:corrfunc})  and isolate the pole contribution of the proton to obtain the hadronic representation of the correlation function:
\begin{equation}
\begin{split}
\Pi^{\rm had}_{\Gamma\Gamma^\prime} (p_p,q) & =  -\frac{m_p}{p_p^2-m_p^2+i \epsilon} \, \lambda_p H_{\Gamma \Gamma^\prime}(p_p,q) \left(\slashed{p}_p + m_p\right) + \ldots \\[2mm]
& = \, P_{\Gamma^\prime} \left(\Pi^{\text{had,} S}_{\Gamma \Gamma^\prime}  + \frac{\slashed{p}_p}{m_p} \, \Pi^{\text{had,}P}_{\Gamma \Gamma^\prime} + \frac{\slashed{q}}{m_p} \, \Pi^{\text{had,}Q}_{\Gamma \Gamma^\prime} + \frac{i \sigma^{p_pq}}{m_p^2} \, \Pi^{\text{had,} T}_{\Gamma \Gamma^\prime} \right) \,,
\end{split}
\label{eq:corrhad}
\end{equation}
with $\epsilon >0$ and infinitesimal, $\sigma^{p q} \equiv \sigma_{\mu\nu} \hspace{0.5mm} p^\mu q^\nu$ with   $\sigma_{\mu \nu} \equiv i/2 \left ( \gamma_\mu \gamma_\nu- \gamma_\nu \gamma_\mu \right)$ and the ellipsis denotes contributions from heavier states,~i.e.~excited states  and the continuum.  The four independent Dirac structures in~\eqref{eq:corrhad} can be used to derive  LCSRs for the form factors~$W^n_{\Gamma \Gamma^\prime} (q^2)$ or combinations of them.  The corresponding scalar functions $\Pi^{\text{had,} \alpha}_{\Gamma \Gamma^\prime} $ depend only on the square of the proton momentum $p_p^2$ and on the square of the momentum transfer~$Q^2 \equiv -q^2$. They are conveniently parametrised in terms of dispersion integrals,
\begin{equation} \label{eq:pispec}
\Pi^{\text{had,}\alpha}_{\Gamma \Gamma^\prime} (p_p^2,Q^2) = \int_{m_p^2}^\infty \! ds \; \frac{\rho^{\text{had,}\alpha}_{\Gamma \Gamma^\prime}(s,Q^2)}{s-p_p^2} \,,
\end{equation}
where  $\alpha = S,P,Q,T$ and we have introduced the spectral densities 
\begin{equation} \label{eq:spectral}
\rho^{\text{had,}\alpha}_{\Gamma \Gamma^\prime}(s,Q^2) \equiv \frac{1}{\pi} \hspace{0.25mm} \im \Pi^{\text{had,}\alpha}_{\Gamma \Gamma^\prime} (s+i \epsilon,Q^2) \,.
\end{equation}

Separating the ground-state contribution from the contribution of heavy states denoted by~$\rho^{\text{cont,}\alpha}_{\Gamma \Gamma^\prime}(s,Q^2)$, the four spectral densities appearing in~(\ref{eq:corrhad}) can be cast into the form
\begin{equation} \label{eq:rhohadalpha}
\rho^{\text{had,}\alpha}_{\Gamma \Gamma^\prime}(s,Q^2) =\, i\lambda_p m_p^2 \, \delta \left(s-m_p^2\right) W^\alpha_{\Gamma \Gamma^\prime}(s,Q^2) + \rho^{\text{cont,}\alpha}_{\Gamma \Gamma^\prime}(s,Q^2) \,, 
\end{equation}
where 
\begin{equation} \label{eq:Walpha} 
\begin{split}  
& \hspace{-5mm} W^S_{\Gamma \Gamma^\prime}(s,Q^2)  = W^0_{\Gamma \Gamma^\prime}(s,Q^2) + \frac{s-Q^2-m_\pi^2}{2 m_p^2} \, W^1_{\Gamma \Gamma^\prime} (s,Q^2) \,, \\[2mm]
W^P_{\Gamma \Gamma^\prime}(s,Q^2) & = W^0_{\Gamma \Gamma^\prime}(s,Q^2)  \,, \qquad W^Q_{\Gamma \Gamma^\prime}(s,Q^2) = W^T_{\Gamma \Gamma^\prime}(s,Q^2)  = W^1_{\Gamma \Gamma^\prime}(s,Q^2)  \,,
\end{split}
\end{equation}
and $m_\pi = 135 \, {\rm MeV}$ is the pion mass. We stress that the relations~(\ref{eq:Walpha})  only hold on-shell,~i.e.~if~$s=m_p^2$. This is however guaranteed by the $\delta \left(s-m_p^2\right) $  function appearing in~(\ref{eq:rhohadalpha}).  Under the  assumption of a global quark-hadron duality~\cite{Poggio:1975af} (see also~\cite{Shifman:2000jv} for a review) the contributions of heavy states can be  approximated~by
\begin{equation}
 \int_{s_0}^\infty ds \, \frac{\rho^{\text{cont,}\alpha}_{\Gamma \Gamma^\prime}(s,Q^2)}{s-p_p^2} \simeq  \int_{s_0}^\infty ds \, \frac{\rho^{\text{\rm QCD,}\alpha}_{\Gamma \Gamma^\prime}(s,Q^2)}{s-p_p^2} \,,
 \label{eq:duality}
\end{equation}
where $\rho^{\text{\rm QCD,}\alpha}_{\Gamma \Gamma^\prime}(s,Q^2)$ are the spectral densities in QCD and we will explain how to  compute them in the next section. The approximation~(\ref{eq:duality}) is expected to work well for a sufficiently large continuum threshold~$s_0$, which is a free parameter and has to be determined within the~LCSR calculation. A more detailed discussion on how to fix~$s_0$ is provided in~Section~\ref{sec:analysis}, but ideally it  is chosen low enough to cover even the lightest excitation, which~is the Roper~resonance with a mass of $1.44 \, {\rm  GeV}$.

\section{LCSR calculation}
 \label{sec:lcsr}

The basic idea of the LCSRs is to derive a result for  $\Pi_{\Gamma\Gamma^\prime} (p_p,q)$ in QCD while parametrising unknown soft contributions in terms of quantities that can be determined by other means.  It can be shown that for large virtualities  $Q^2 \gg \Lambda_{\rm QCD}^2$ and $P_p^2 \equiv -p_p^2 \gg \Lambda_{\rm QCD}^2$ with $\Lambda_{\rm QCD} \simeq 300 \, {\rm  MeV}$ the QCD scale,  the integrand of the correlator~\eqref{eq:corrfunc} can be approximated by an expansion on the light-cone $x^2 \sim 1/Q^2 \simeq 0$~(see~\cite{Colangelo:2000dp} and references therein). Schematically, this light-cone expansion takes the form
\begin{equation}
T \left[ Q_{\Gamma \Gamma^\prime}(x) \bar{\eta}(0) \right] = \sum_k C_k(x) \mathcal{O}_k(0) \,,
\label{eq:twistexp}
\end{equation}
where  the Wilson coefficients $C_k$ encode the hard scattering process and the objects $\mathcal{O}_k$ are composite operators of twist $k$. The matrix elements of these composite operators correspond to the light-cone DAs of the pion  which are non-perturbative objects. Performing a Borel transformation with respect to  $P_p^2$ then yields an expansion in inverse powers of the two scales that enter our calculation,~i.e.~it leads to a power expansion in $\Lambda_{\rm QCD}^2/M^2$ and~$\Lambda_{\rm QCD}^2/Q^2$, where $M$ denotes the Borel mass associated to $P_p^2$ $\big($cf.~\eqref{eq:boreltrafo}$\big)$. In our article we will provide explicit  LCSR expressions that include the leading contributions in this expansion, namely the twist-2 and twist-3 DAs. We will however also comment on the possible impact of twist-4 contributions in all cases where such terms could be phenomenologically relevant (cf.~Section~\ref{sec:analysis}).

In order to carry out the light-cone expansion, we need to choose an explicit form for the proton current $\eta$. The most general choice with the appropriate quantum numbers (at lowest order in derivatives and spin) can be written as a linear combination of the following two currents~\cite{Ioffe:1982ce}:
\begin{equation}
\eta_1(x) = 2 \epsilon^{abc} \left(u_a^T(x) C \gamma_5 d_b(x)\right)  u_c(x)  \,, \qquad 
\eta_2(x) = 2 \epsilon^{abc} \left(u_a^T(x) C d_b(x)\right) \gamma_5 u_c(x) \,. 
\end{equation}
The current $\eta_1$ excites the ground state as well as heavier states, while $\eta_2$ almost exclusively excites heavier states~\cite{Leinweber:1994nm}. As a result the coupling strength of $\eta_1$ $\big($cf.~\eqref{eq:protoncoupl}$\big)$ to the proton state is larger by a factor of about $100$ than that of $\eta_2$. Due to its weak coupling to the proton state, the contribution of the current $\eta_2$ is expected to be very small in the case at hand, and we therefore choose for simplicity
\begin{equation} \label{eq:ourprotoncurrent}
\eta(x) \equiv \eta_1(x) \,,
\end{equation}
neglecting a possible admixture of $\eta_2$. Notice that our choice of proton current corresponds to the interpolator usually used in LQCD calculations.

The expansion of the time-ordered product that occurs in the twist expansion~\eqref{eq:twistexp} is carried out by partially contracting the quark fields,
\begin{equation} \label{eq:ope}
\begin{split}
T \left[Q_{\Gamma \Gamma^\prime}(x) \bar{\eta} (0) \right] =& -\frac{1}{2}\,  \epsilon_{ijk} \, \epsilon_{abc} \, P_{\, \Gamma^\prime} \, \bigg\{ \left(\bar{u}^a(0) \Gamma_A u^i(x) \right) \\[2mm]
& \hspace{-1cm} \times \Big[  S_u^{kc}(x) \gamma_5 \tilde{S}_d^{jb}(x) P_{\, \Gamma} \Gamma^A  + S_u^{kc}(x) \Tr \left( \Gamma^A \gamma_5 \tilde{S}_d^{jb}(x) P_{\, \Gamma} \right) \\[2mm]
& \hspace{-0.5cm} +\Gamma^A \gamma_5 \tilde{S}_d^{jb}(x) P_{\, \Gamma} S_u^{kc}(x)   + \Gamma^A \Tr \left( S_u^{kc}(x) \gamma_5 \tilde{S}_d^{jb}(x) P_{\, \Gamma} \right) \Big] \\[2mm]
& \hspace{-1.5cm}  + \left(\bar{d}^a(0) \Gamma_A d^i(x) \right)  \Big[  S_u^{kc}(x) \gamma_5 \tilde{\Gamma}^{A} P_{\, \Gamma} S_u^{jb}(x)  + S_u^{kc}(x) \Tr \left(S_u^{jb}(x) \gamma_5 \tilde{\Gamma}^{A} P_{\, \Gamma} \right) \Big]  \bigg\} \,.
\end{split}
\end{equation}
Here we have employed the  following basis of gamma matrices
\begin{equation} \label{eq:basis}
\Gamma_A = \left\{ 1, \gamma_5, \gamma^\rho, i \gamma^\rho \gamma_5 , \frac{1}{\sqrt{2}} \sigma^{\rho \sigma} \right\} \,,
\end{equation}
used the notation $\tilde{\Gamma}_A \equiv C \hspace{0.25mm} \Gamma_A^T \hspace{0.25mm}  C$ with $C= i \gamma^2 \gamma^0$ and a summation over the index $A$ is  implicit. The pairwise contraction of up (down) quark fields is denoted by $S^{ij}_{u}(x)$ $\big($$S^{ij}_{d}(x)$$\big)$, $i,j,k$ are colour indices and ${\rm Tr}$ denotes a trace over  Dirac matrices. Hereafter we will work in  the isospin limit and will therefore drop the flavour index of the contraction. 

\begin{figure}[t]
\centering
\includegraphics[scale=.8]{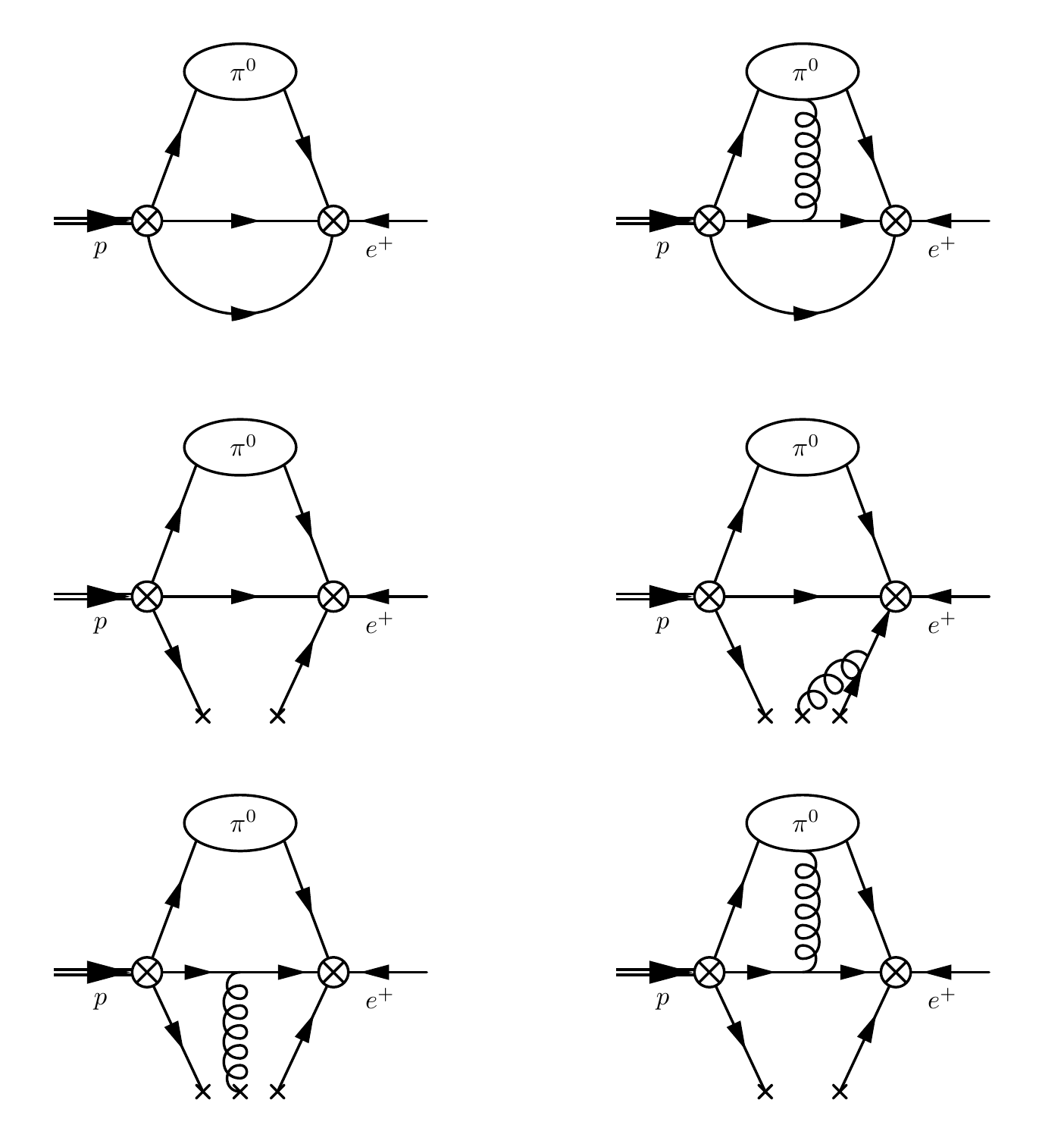}
\caption{Feynman diagrams contributing to the light-cone expansion of~\eqref{eq:twistexp} at the twist-2~and twist-3 level including factorised higher-twist contributions.  The two vertices with a circled cross denote insertions of the currents $Q_{\Gamma \Gamma^\prime}(x)$ and~$\eta(0)$. We attach the external proton and positron lines for illustration even though they do not enter the LCSR computation.  The diagrams shown in the top row  result from the light-cone expansion~(\ref{eq:quarkprop}). The diagrams in the middle and bottom row instead originate from factorised higher-twist contributions, which involve the condensates $\braket{\bar{q}q}$ and $\braket{\bar{q}g_s G \cdot \sigma q}$ of~(\ref{eq:quarkproplocal})~and~(\ref{eq:mixedcond}) (crosses on the bottom of the diagram). See text for further details.}
\label{fig:qcddia}
\end{figure}

With the help of~(\ref{eq:basis}) it is possible to derive the following completeness relation:
\begin{equation}
u(x) \bar{u}(0) = -\frac{1}{4} \left(\bar{u}(0) \Gamma_A u(x) \right) \Gamma^A \,.
\end{equation}
The contracted fields need to be expanded for light-like distances (including single-gluon emission), which reads~\cite{Balitsky:1987bk}
\begin{equation}
S^{ij}(x) =\frac{i\slashed{x}}{2\pi^2 x^4} \, \delta^{ij} - \frac{ig_s}{16\pi^2 x^2} \int_0^1 \! du\, G^{ij}_{\mu\nu}(ux) \left[ \bar u \hspace{0.25mm} \slashed{x} \sigma^{\mu\nu} + u  \hspace{0.25mm} \sigma^{\mu\nu} \slashed{x} \right] + \ldots \,,
\label{eq:quarkprop}
\end{equation}
where the ellipsis represents terms that lead to contributions of twist higher than three, $g_s$ is the QCD coupling constant, we have employed the short-hand notation  $G_{\mu\nu}^{ij} \equiv  G^A_{\mu\nu} \hspace{0.25mm} T_A^{ij}$  for the gluon field strength tensor with $T_A^{ij}$ the $SU(3)$ generators and defined $\bar{u} \equiv 1-u$. We neglect contributions proportional to the quark masses because they are numerically negligible. In the following we consider only single-gluon interactions which is consistent with truncating the expansion~(\ref{eq:twistexp}) after the leading-twist contribution~\cite{Braun:1989iv}.  This leads to the two types of one-loop diagrams that are displayed in  the top row of Figure~\ref{fig:qcddia}. 

In addition to these  leading-order terms factorised contributions of higher twist and multiplicity turn out to be numerically relevant in the case at hand. Such contributions originate from operators with four quark fields or four quark fields and one gluon, such that only one pair of quarks is contracted in the time-ordered product. Part of the respective amplitudes can be approximated by a factorisation into two- or three-particle DAs (of twist two and three) and vacuum condensates of the remaining quark and gluon fields. Such contributions scale with a smaller power of $1/Q^2$ than genuine, non-factorisable terms of higher twist and instead are suppressed by powers of $1/M^2$~\cite{Braun:1999uj,Agaev:2010aq}. Effectively, these factorised contributions can be taken into account by replacing one of the contractions~$S^{ij}(x)$ in the expression~(\ref{eq:ope}) by the appropriate local terms and condensates  as encoded by~\cite{Leinweber:1995fn}
\begin{equation}
 \Delta S^{ij}(x) =  -\frac{\braket{\bar{q}q}}{12} \, \delta^{ij} \left(1+\frac{m_0^2 \hspace{0.25mm} x^2}{16}\right) - \frac{ig_s}{32\pi^2 x^2} \, G^{ij}_{\mu\nu}(0) \left[ \slashed{x} \sigma^{\mu\nu}+ \sigma^{\mu\nu}\slashed{x} \right] + \ldots \,.
\label{eq:quarkproplocal}
\end{equation}
Here the ellipsis denotes higher-dimensional condensates and terms with additional gluons, which are neglected  in our work because they are numerically small. The parameter~$m_0$ entering~(\ref{eq:quarkproplocal}) is associated with the mixed condensate 
\begin{equation} \label{eq:mixedcond}
\braket{\bar{q}g_s G \cdot \sigma q} = m_0^2 \braket{\bar{q}q} \,,
\end{equation}
where $G \cdot \sigma \equiv G_{\mu\nu} \hspace{0.25mm} \sigma^{\mu \nu}$. The diagrams resulting from the local expansion~(\ref{eq:quarkproplocal}) of the contraction are displayed in the middle and bottom row of Figure~\ref{fig:qcddia}.

The uncontracted quark bilinears in~\eqref{eq:ope} form a pion and still need to be expanded around light-like distances to obtain the light-cone DAs. The pion~DAs have been extensively studied in  the literature~(see~\cite{Khodjamirian:2020mlb} for a state-of-the-art discussion), and they have definite twist.  The only  twist-2 pion  DA is given by (cf.~for instance~\cite{Ball:1998je,Ball:2004ye})
\begin{equation}
\braket{\pi^0(p_\pi)| \hspace{0.25mm}  \bar{q}(0) \gamma^\mu\gamma_5 \tau^3 q (x) \hspace{0.25mm} |0}  = -\frac{if_\pi}{ \sqrt{2}}\, p_\pi^\mu \int_0^1 \! du \, e^{i \bar{u} p_\pi x} \, \phi^{(2)}(u,\mu) \,,
\label{eq:DA2}
\end{equation}
where  $q \equiv (u \; d)^T$ and $\tau^3 \equiv   \sigma^3/2$ with $\sigma^3 = \text{diag} \left (1,-1 \right )$ the third Pauli matrix, while~$f_\pi$  denotes the pion decay constant given by~$f_\pi = \left ( 130.2 \pm 0.8 \right ) \, {\rm  MeV}$~\cite{Aoki:2019cca}.\footnote{We employ the Fock-Schwinger gauge,~i.e.~$x^\mu G^A_\mu = 0$ with $G^A_\mu$ the gluon field, such that the Wilson lines which enter the definition of the DAs are equal to 1.} The parameters~$u$ and~$\bar{u}$ correspond to the momentum fractions of the two quarks that form the pion.  The~renormalisation scale~$\mu$ that appears in the twist-2 pion  DA $\phi^{(2)}(u,\mu) $ is set equal to~$1 \, {\rm GeV}$ for most of this work. Higher-order contributions to the matrix element~\eqref{eq:DA2} arise at the twist-4 level.  There are two two-particle  twist-3  DAs called $ \phi^{(3)}_p(u,\mu)$ and~$\phi^{(3)}_\sigma(u,\mu)$. These   are defined by~\cite{Ball:1998je,Ball:2004ye}
\begin{align}
\braket{\pi^0(p_\pi)| \hspace{0.25mm}  \bar{q}(0) i\gamma_5 \tau^3 q (x) \hspace{0.25mm}  |0} & = \frac{f_\pi \mu_\pi}{\sqrt{2}} \int_0^1 \! du \, e^{i \bar{u} p_\pi x}  \, \phi^{(3)}_p(u,\mu)\,, \label{eq:DAp} \\[2mm]
\braket{\pi^0(p_\pi)| \hspace{0.25mm}  \bar{q}(0) \sigma^{\mu\nu}\gamma_5 \tau^3 q (x)  \hspace{0.25mm}  |0} & = -\frac{i f_\pi \mu_\pi}{6 \sqrt{2}}\left(1-\rho_\pi^2\right) \left(p_\pi^\mu x^\nu-p_\pi^\nu x^\mu\right) \notag \\[-2.5mm]  \label{eq:DAs} \\[-2.5mm] 
& \phantom{xx} \times \int_0^1 \! du \, e^{i \bar{u} p_\pi x} \, \phi^{(3)}_\sigma(u,\mu) \,. \notag 
\end{align}

We also include the only twist-3 three-particle DA  called $\mathcal{T}^{(3)}(\alpha_d,\alpha_u,\alpha_g,\mu)$, which depends on the  momentum fractions $\alpha_d$, $\alpha_u$ and $\alpha_g$ of the down quark, up quark and gluon, respectively, as well as on $\mu$. This object is defined as follows~\cite{Ball:1998je,Ball:2004ye}
\begin{align}
&\braket{\pi^0(p_\pi)| \hspace{0.25mm} \bar{q}(0) \sigma^{\mu\nu}\gamma_5 g_s G^{\alpha \beta}(ux)\tau^3 q (x)  \hspace{0.25mm}  |0} = \notag \\[2mm]
& \phantom{xx}  \frac{i f_\pi \mu_\pi}{\sqrt{2} } \left(p_\pi^\alpha p_\pi^\mu g^{\nu\beta}-p_\pi^\alpha p_\pi^\nu g^{\mu\beta} + p_\pi^\beta p_\pi^\nu g^{\alpha\mu}-p_\pi^\beta p_\pi^\mu g^{\alpha \nu}\right)   \label{eq:DA3p} \\[2mm]
&  \phantom{xx}  \quad \times \int_0^1 \! d\alpha_d \hspace{0.5mm}  d\alpha_u \hspace{0.5mm} d \alpha_g \hspace{0.5mm} \delta(1-\alpha_d-\alpha_u -\alpha_g)  \hspace{0.5mm} e^{i \left (\alpha_u + u \alpha_g \right ) p_\pi x } \, \mathcal{T}^{(3)}(\alpha_d,\alpha_u,\alpha_g,\mu) \,.  \notag
\end{align}

The  normalisation of the twist-3 DAs contains the sum of the up- and down-quark mass, which fixes the values of the parameters $\mu_\pi$ and $\rho_\pi$ as well as the quark condensate via the  Gell-Mann--Oakes--Renner~(GMOR) relation $m_\pi^2 \simeq -2 \left (m_u + m_d \right ) \braket{\bar{q}q}/f_\pi^2$ first derived in the article~\cite{GellMann:1968rz}. One obtains
\begin{equation}
\mu_\pi \equiv \frac{m_\pi^2}{m_u+m_d} \simeq - \frac{2 \braket{\bar{q}q}}{f_\pi^2} \,, \qquad \rho_\pi \equiv \frac{m_u+m_d}{m_\pi} \simeq - \frac{f_\pi^2 \hspace{0.25mm} m_\pi}{2 \braket{\bar{q}q}} \,.
\label{eq:GMOR}
\end{equation}
The DAs can be obtained by a conformal expansion~\cite{Ball:1998je}. Explicit formulas for the DAs appearing in our work are provided in Appendix~\ref{app:DA}.

Summing up all the contributions of Figure~\ref{fig:qcddia}, using~\eqref{eq:DA2} to~\eqref{eq:DA3p} and performing a Fourier integration allows us one derive an analytic expression for the QCD correlation function
\begin{equation}
\Pi^{\rm QCD}_{\Gamma\Gamma^\prime} (p_p,q) = P_{\Gamma^\prime} \left(\Pi^{\text{QCD,}S}_{\Gamma \Gamma^\prime} + \frac{\slashed{p}_p}{m_p} \, \Pi^{\text{QCD,}P}_{\Gamma \Gamma^\prime} + \frac{\slashed{q}}{m_p}\,  \Pi^{\text{QCD,}Q}_{\Gamma \Gamma^\prime} + \frac{i \sigma^{p_pq}}{m_p^2} \, \Pi^{\text{QCD,}T}_{\Gamma \Gamma^\prime} \right) \,.
\label{eq:corrQCD}
\end{equation}
The explicit expressions for the analytic results of $\Pi^{\text{QCD,}\alpha}_{\Gamma \Gamma^\prime} \left(p_p^2,Q^2\right) $ are somewhat lengthy and therefore provided in Appendix~\ref{app:formulaeLCSRs}. For the matching with the hadronic representation~(\ref{eq:corrhad}), we also introduce QCD  spectral densities like it has been done in~(\ref{eq:pispec}) and~(\ref{eq:spectral}) for the hadronic case. The matching conditions for the  LCSRs then read
\begin{equation}
\Pi^{\text{had,}\alpha}_{\Gamma \Gamma^\prime} \left(p_p^2,Q^2\right)\,  \overset{!}{=} \, \Pi^{\text{QCD,}\alpha}_{\Gamma \Gamma^\prime} \left(p_p^2,Q^2\right) \,.
\label{eq:opematching}
\end{equation}

Using quark-hadron duality in the form~\eqref{eq:duality}, we can subtract the unknown contributions of heavy states from the  LCSRs. Effectively, this procedure cuts off the spectral integral computed  in~QCD  at the continuum threshold~$s_0$.  Applying a Borel transformation with respect to $P_p^2$ to both sides of the sum rules suppresses heavy contributions exponentially and generically improves the  accuracy of the LCSR approach --- the accuracy of our~LCSRs will be  investigated in~Section~\ref{sec:analysis}. The Borel transformations also remove all terms that are polynomial in $P_p^2$, which sets all divergent contributions of the dispersion integrals as well as the ultraviolate~(UV) divergences of the diagrams in Figure~\ref{fig:qcddia} to zero. The Borel transforms that enter our~LCSR analysis as well as other calculational details are given in~Appendix~\ref{app:formulae}. After Borel transformation the matching conditions~(\ref{eq:opematching})  take the form
\begin{equation} \label{eq:ourLCSRs} 
i\lambda_p m_p^2 \, e^{-\frac{m_p^2}{M^2}} \, W^\alpha_{\Gamma \Gamma^\prime}(s_0, Q^2)  = \int^{s_0}_0 \! ds\, e^{-\frac{s}{M^2}} \, \rho^{\text{QCD,}\alpha}_{\Gamma \Gamma^\prime} \left(s,Q^2\right) \,,
\end{equation}
where  the expressions for the form factors $W^\alpha_{\Gamma \Gamma^\prime}(s_0, Q^2)$ can be found in~(\ref{eq:Walpha}). By an appropriate combination of  the four independent relations~(\ref{eq:ourLCSRs}) one can derive two LCSRs for each of the two form factors appearing in~(\ref{eq:W0W1}). Hereafter we will refer to these combinations as $W^{0, P}_{\Gamma \Gamma^\prime}(s_0, Q^2)$, $W^{0, S+T}_{\Gamma \Gamma^\prime}(s_0, Q^2)$, $W^{1, Q}_{\Gamma \Gamma^\prime}(s_0, Q^2)$ and $W^{1, T}_{\Gamma \Gamma^\prime}(s_0, Q^2)$. Notice that the~LCSRs~(\ref{eq:ourLCSRs}) depend on  two unphysical parameters, namely the continuum threshold~$s_0$ and the Borel mass $M$. However, the results of a LCSR calculation can only be trusted if  the final predictions are to a certain extent independent of the exact choice of~$s_0$~and~$M$. In~Section~\ref{sec:analysis} we will provide criteria that allow to assess the convergence properties of~(\ref{eq:ourLCSRs}), which we will then us to estimate the uncertainties that plague our~LCSR results for the form factors~$W^{n,\alpha}_{\Gamma \Gamma^\prime} (Q^2)$.

The coupling $\lambda_p$ in~(\ref{eq:ourLCSRs}) is in principal known from LQCD calculations (see~\cite{Leinweber:1995fn} and references therein), but it can as well be extracted from local QCD sum rules of the two-point correlator
\begin{equation} \label{eq:lampsr} 
i \int \!d^4x \, e^{ipx} \braket{0| \, T \left[  \eta(x) \bar{\eta}(0) \hspace{0.125mm} \right]  |0} = -\lambda_p^2\; \frac{\slashed{p} + m_p}{p^2-m_p^2+i \epsilon} + \ldots \,,
\end{equation}
where the ellipsis denotes the contributions of heavier states.  Using $\lambda_p$ from sum rules has the salient advantage that in this way the  uncertainties of the form factors due to the input parameters such as the quark condensate $\braket{\bar{q}q}$ or $m_0^2$ are reduced.\footnote{In $B$-meson to light-meson transitions this procedure even leads to a partial cancellation of perturbative corrections, which improves the convergence of the sum rules~\cite{Ball:2004ye}.} We therefore choose to~fix $\lambda_p$ with the help of sum-rule techniques rather than to take $\lambda_p$ from LQCD computations. The sum rule derived from the structure $\slashed{p}$ of the  correlator~(\ref{eq:lampsr}) is typically disregarded due to uncontrollably large radiative corrections as well as large contributions from heavier states~\cite{Leinweber:1995fn}. We thus extract~$\lambda_p$ from the sum rule for the mass term. The~QCD result for the sum rule can be derived by performing a local OPE around $x \simeq 0$ of the time-ordered product in~(\ref{eq:lampsr}). If the momentum flow through the correlator is deeply space-like, i.e.~$-p^2 \gg \Lambda_{\rm QCD}^2$, this leads to a convergent expansion of local operators with increasing mass dimension~\cite{Wilson:1969zs}. By plugging the OPE into~(\ref{eq:lampsr}) one then obtains an expansion of the correlator in terms of condensates. In this work, contributions including condensates up to dimension seven are included but no perturbative QCD corrections.  Using~the proton current~(\ref{eq:ourprotoncurrent}) yields~\cite{Leinweber:1995fn}
\begin{equation}  
\lambda_p^2 = -\frac{\braket{\bar{q}q}}{16 \hspace{0.125mm} \pi^2 \hspace{0.25mm} m_p^3}  \, e^{\frac{m_p^2}{\overbar{M}^2}} \left [ \, 7 \overbar{M}^4 E_2 \left( \frac{\bar{s}_0}{\overbar{M}^2} \right) - 3 \hspace{0.25mm} m_0^2  \hspace{0.25mm} \overbar{M}^2 E_1 \left( \frac{\bar{s}_0}{\overbar{M}^2} \right)  +  \frac{19 \hspace{0.25mm} \pi^2}{18} \left \langle  \frac{\alpha_s}{\pi} \hspace{0.25mm} G^2 \right \rangle  \, \right ] \,, 
\label{eq:localsr}
\end{equation}
with $\alpha_s \equiv g_s^2/(4 \pi)$, $G^2 \equiv G_{\mu \nu}^A G^{A, \mu \nu}$ and 
\begin{equation} \label{eq:Enx}
E_n (x) \equiv 1- e^{-x} \sum_{k=0}^{n-1} \frac{x^k}{k!} \,.
\end{equation}
The parameters $\bar{s}_0$  and $\overbar{M}$ denote the continuum threshold  and the Borel mass of the local sum rule~(\ref{eq:localsr}).  These parameters can be related to the corresponding parameters of the~LCSRs, because the Borel mass is connected to the momentum flow through the proton current.  However, we assume for simplicity that $\bar{s}_0$ and $\overbar{M}$  are independent parameters and determine them such that the value of $\lambda_p$ does not depend too strongly on the specific~choice.  

Notice finally that the sign of $\lambda_p$ is not fixed by~(\ref{eq:localsr}). More generally, the  sign of $\lambda_p$ depends on the (unphysical) phase of the nucleon wave function. The same holds for the overall sign of the form factors $W^{n, \alpha}_{\Gamma \Gamma^\prime} (s_0,Q^2)$ that are determined from~(\ref{eq:ourLCSRs}). The relative sign between~$W^{0}_{\Gamma \Gamma^\prime} (Q^2)$  and $W^{1}_{\Gamma \Gamma^\prime} (Q^2)$ is however fixed by our sum rules. In the following, we will choose a negative sign for the coupling strength of the proton current,~i.e.~we will employ~$\lambda_p <0$.

\section{Numerical analysis}
\label{sec:analysis} 

To derive physical predictions from~\eqref{eq:ourLCSRs}, we need to find regions where the LCSRs converge sufficiently fast as an expansion in $\Lambda_{\rm QCD}^2/Q^2$ and $\Lambda_{\rm QCD}^2/M^2$ and where the sum rules are to a certain extent insensitive to the choice of the continuum threshold $s_0$ and the Borel mass $M$. Therefore, the Borel mass $M$ has to be chosen well above the QCD scale $\Lambda_{\rm QCD}$ but at the same time well below the mass of the lightest excitation. These conditions are formulated more precisely in the following, and they lead to a set of requirements which are then applied to each of the four LCSRs~\eqref{eq:ourLCSRs} as well as the  local sum rule~(\ref{eq:localsr}).

In order to eliminate contributions other than the proton in our sum-rule calculations~\cite{Shifman:1978bx}, we use $s_0= \bar{s}_0 =(1.44 \, \text{GeV})^2$ as a central value in all cases. This value of the continuum threshold~   $s_0$ corresponds to the mass of the lightest excited state in the nucleon spectrum,~i.e.~the Roper resonance. We then vary~$s_0$ (and $\bar{s}_0$) between $(1.4 \, \text{GeV})^2$ and~$(1.5 \, \text{GeV})^2$ to estimate the uncertainty related to the choice of the continuum threshold. A~similar procedure has been adopted in~\cite{Braun:2001tj,Braun:2006hz}, and our  choice can be further  motivated by the observation that for values in this interval, the sum rule~(\ref{eq:localsr}) leads to a good agreement with the LQCD results for $\lambda_p$ (see for instance~\cite{Gavela:1988cp,Chu:1993cn,Leinweber:1994gt,Bali:2019ecy}). 

A lower bound on  $M$ is determined by demanding sufficient suppression of higher powers in the OPE. In particular, we require that the contribution of the highest dimensional condensate in each LCSR does not amount to more than approximately $30\%$ of the total QCD result. An upper limit on  $M$ is instead obtained by demanding that the ground-state contribution in the hadronic representation constitutes at least $50\%$ of the dispersion integral. In other words the contributions of the heavy states, which we model by the QCD result, are smaller or equal than approximately  $50\%$ of the total result,
\begin{equation}
\dfrac{\left|  P^{\text{QCD},\alpha}_{\Gamma\Gamma^\prime}  (s_0, \infty) \right| }{\left|  P^{\text{QCD},\alpha}_{\Gamma\Gamma^\prime}  (0, \infty)   \right|} \lesssim 0.5 \,,
\end{equation}
with
\begin{equation}
P^{\text{QCD},\alpha}_{\Gamma\Gamma^\prime} (s_1, s_2)=  \int_{s_1}^{s_2} \! ds \; \dfrac{\rho^{\text{QCD},\alpha}_{\Gamma\Gamma^\prime}\left(s,Q^2\right)}{s-p_p^2} \,.
\end{equation}
We then vary the Borel mass $M$ in this so obtained Borel window to estimate the systematic uncertainty related to the variation of this unphysical parameter. 

\begin{figure}[t]
\centering
\includegraphics[width=\textwidth]{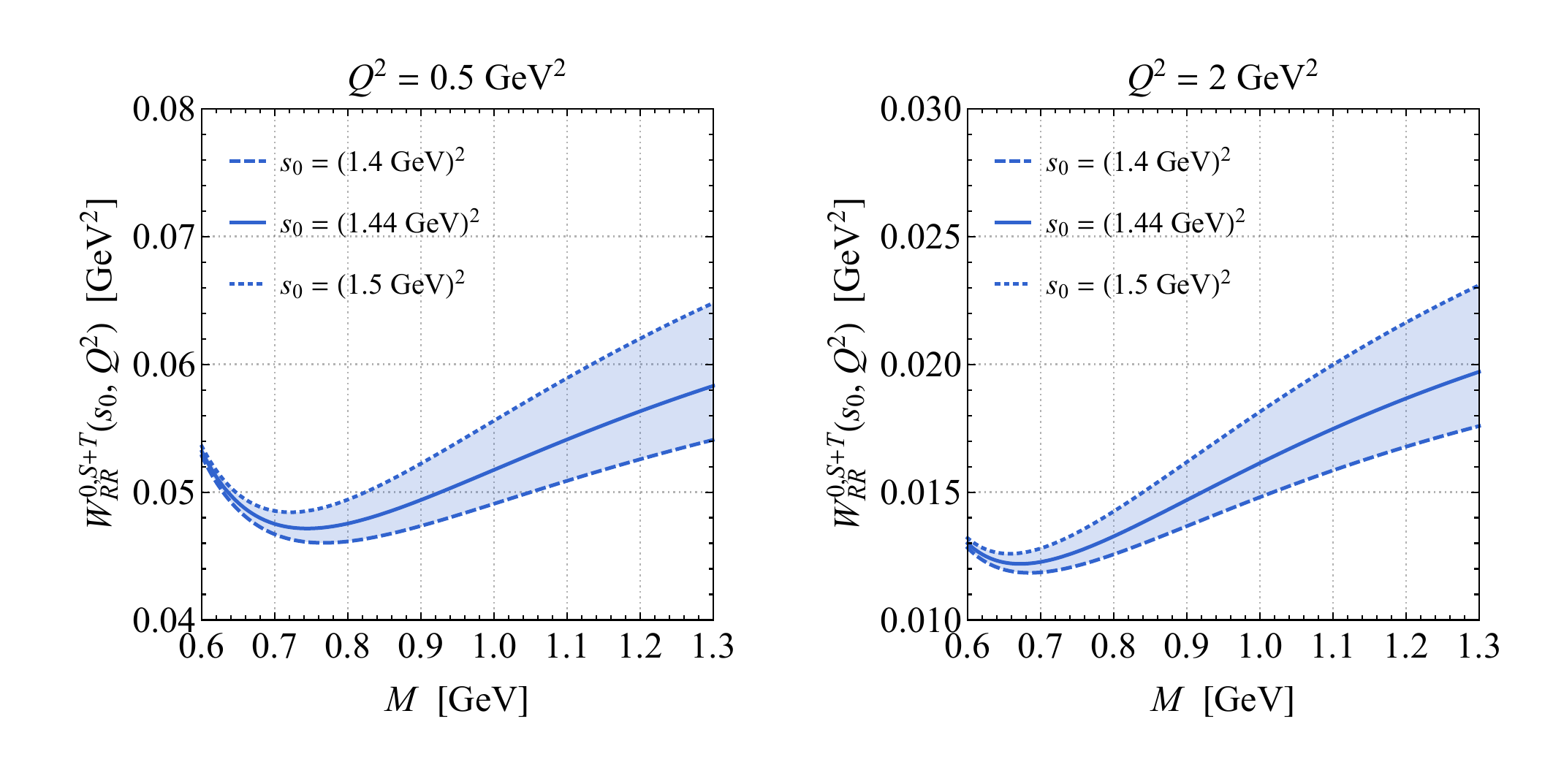}
\vspace{-10mm}
\caption{Form factor  $W^{0, S+T}_{RR} (s_0, Q^2)$ as a function of the Borel mass $M$ for three values of the continuum threshold $s_0$. The left (right) plot shows the results at $Q^2 = 0.5 \, {\rm GeV}^2$~$\big($$Q^2 = 2 \, {\rm GeV}^2$$\big)$.}
\label{fig:borel1}
\end{figure}

The physical values of the form factors do not depend on the choice of the continuum threshold $s_0$ or the Borel mass~$M$. The residual dependence of the form factors extracted from the LCSRs on these parameters originates from the truncation of the expansion in~$\Lambda_{\rm QCD}^2/M^2$ at a finite order and the effective description of the a priori unknown contributions of heavy states. Therefore, the predictions of the sum rules are reliable if the dependence on the unphysical parameters is weak, and thus the uncertainties related to the variation of these parameters also quantifies the validity of the predictions.

\begin{figure}[t]
\centering
\includegraphics[width=\textwidth]{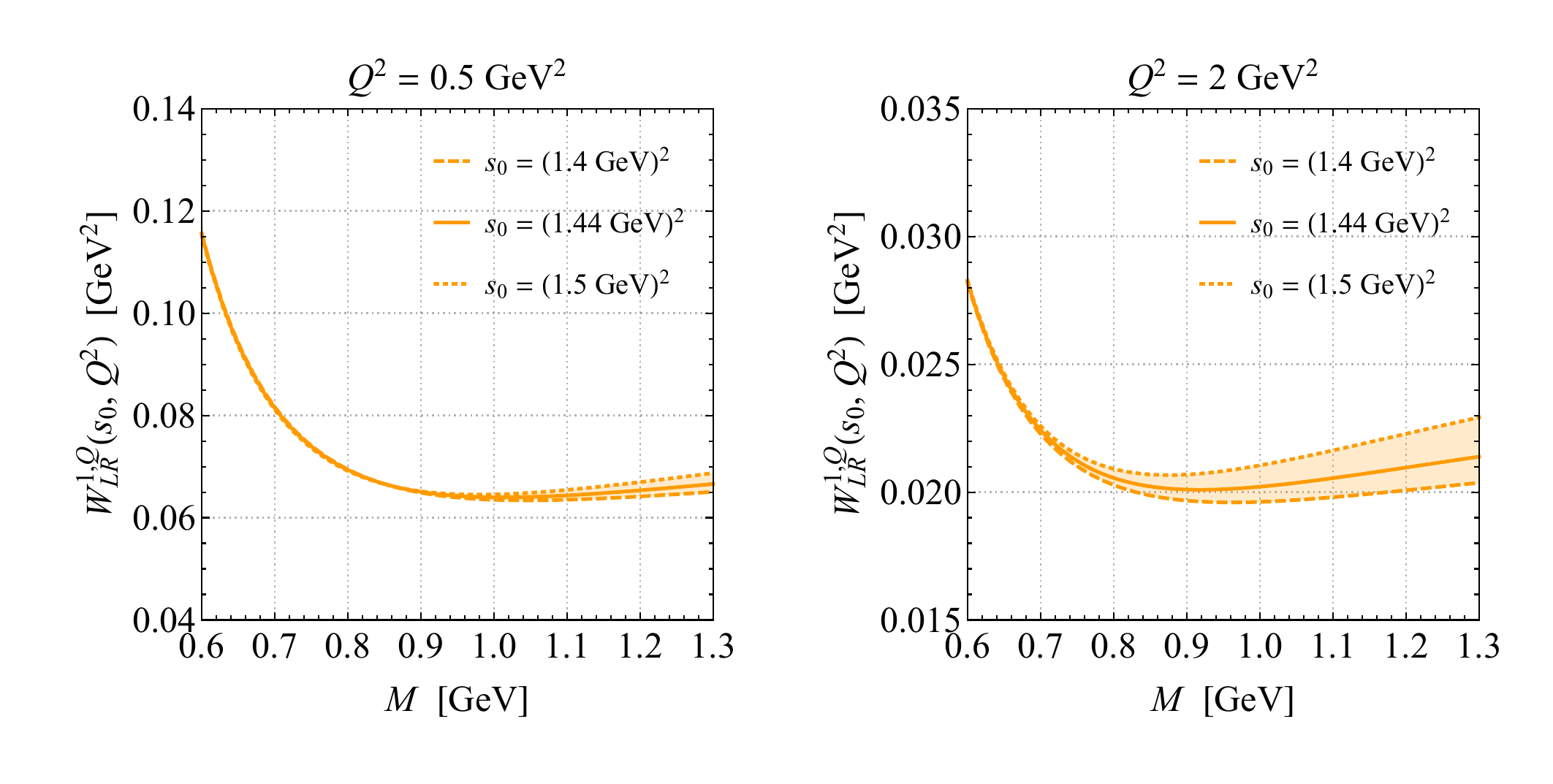}
\vspace{-10mm}
\caption{As Figure~\ref{fig:borel1} but for the form factor $W^{1, Q}_{LR} (s_0, Q^2)$. }
\label{fig:borel2}
\end{figure}

In order to illustrate  the latter statements we  show in Figure~\ref{fig:borel1} and Figure~\ref{fig:borel2} the dependence of the form factor $W^{0, S+T}_{RR} (s_0, Q^2)$  and $W^{1, Q}_{LR} (s_0, Q^2)$  on the Borel mass~$M$, respectively. In each panel predictions are displayed for the following three different values of the continuum threshold $s_0 = (1.4 \, {\rm GeV})^2$ (dashed lines), $s_0 = (1.44 \, {\rm GeV})^2$~(solid lines) and $s_0 = (1.5 \, {\rm GeV})^2$ (dotted lines),  and each figure contains results where the form factors are evaluated at  $Q^2 = 0.5 \, { \rm GeV}^2$ (left panels) and $Q^2 = 2 \, { \rm GeV}^2$~(right panels). The shown predictions have been obtained for the central values of the input parameters as given in~(\ref{eq:qbqcond})~to~(\ref{eq:m0gluon2}) and~(\ref{eq:a2a4})~to~(\ref{eq:eta3}). One sees that for very small values of $M$ the form factors steeply increase because the power suppression in $1/M^2$ becomes ineffective. On the other hand, for large values of $M$ the exponential suppression of heavier states due to the factor $e^{-s/M^2}$ in the dispersion integrals~\eqref{eq:ourLCSRs} is not present. This in turn leads to a stronger sensitivity on $M$ such that the form factors increase again for larger Borel masses. Also the sensitivity to the continuum threshold is more pronounced  if~$M$  gets closer to $s_0$,  as indicated by the widening  of the coloured bands in Figures~\ref{fig:borel1} and~\ref{fig:borel2}.  One~can furthermore observe that the sensitivity on the unphysical parameters becomes stronger for larger values of $Q^2$. We remark that  for $Q^2 \gtrsim 2 \, {\rm GeV}$ this effect saturates such that the plots on the right-hand side of Figures~\ref{fig:borel1} and~\ref{fig:borel2} represent in a sense  worst-case scenarios. The results of the other LCSRs  behave similarly to $W^{0, S+T}_{RR} (s_0, Q^2)$  and $W^{1, Q}_{LR} (s_0, Q^2)$,   so we do not show their dependence on $M$ explicitly.

By considering the momentum range $0.5 \, { \rm GeV}^2 \leq Q^2 \leq 2.5 \, { \rm GeV}^2$, we find the Borel~windows $0.7 \, \text{GeV} \leq M \leq 1.1 \, \text{GeV}$ for the LCSRs with $\alpha=S,Q,T$ and  $\Gamma \Gamma^\prime= RR$ and $0.7 \, \text{GeV} \leq M \leq 1 \, \text{GeV}$ for  the LCSRs with $\alpha=S,Q,T$ and $\Gamma \Gamma^\prime= LR$. The LCSRs~for the structure $\alpha=P$ do not meet the above requirements in the $Q^2$ region of interest, because the contributions of heavy states are large and even dominate the sum rules for certain values of $Q^2$. In order to have one common Borel window for each chirality combination, we also choose in the case $\alpha=P$ either  $0.7 \, \text{GeV} \leq M \leq 1.1 \, \text{GeV}$  or  $0.7 \, \text{GeV} \leq M \leq 1 \, \text{GeV}$ as our Borel window when studying the $Q^2$ dependence of $W^{0,P}_{\Gamma \Gamma^\prime} (Q^2)$.

In our numerical analysis of the LCSRs we use $(m_u+m_d)/2 = (3.410 \pm 0.043) \, {\rm MeV}$~\cite{Aoki:2019cca} which corresponds to the $\overline{\rm MS}$ value at $2 \, {\rm GeV}$.  Using the two-loop renormalisation group~(RG) running and the one-loop threshold corrections as implemented in {\tt RunDec}~\cite{Chetyrkin:2000yt,Herren:2017osy}, we obtain at~$1 \, {\rm GeV}$ the value $m_u+m_d = (8.60 \pm 0.11 ) \, {\rm MeV}$. Employing the GMOR relation this value leads to 
\begin{equation} \label{eq:qbqcond}
\braket{\bar{q}q}  = - \big ( \left ( 256 \pm 2 \right ) \, {\rm MeV} \big ) ^3\,, 
\end{equation}
if the leading-order chiral corrections of~\cite{Bordes:2010wy} are included and uncertainties are added in quadrature. For the non-perturbative parameters defined in~(\ref{eq:GMOR}) we then find  
\begin{equation} \label{eq:mupirhopi}
\mu_\pi = \left (1.98 \pm 0.05 \right )  {\rm GeV} \,, \qquad 
\rho_\pi = 0.068 \pm  0.002 \,. 
\end{equation}
The parameter $m_0$ for the mixed condensate as well as the pure-gluon condensate are known from sum-rule estimates evaluated at $1 \, {\rm GeV}$. We will use the values and uncertainties from~\cite{Ioffe:2002ee} which are widely accepted. The relevant numbers read
\begin{equation} \label{eq:m0gluon2}
m_0^2 = \left ( 0.8 \pm 0.2 \right ) \text{GeV}^2\,,\qquad  
\left \langle  \frac{\alpha_s}{\pi} \hspace{0.25mm} G^2 \right \rangle = \left ( 0.009 \pm 0.009 \right )  \text{GeV}^4\,. 
\end{equation}
We remark that using~(\ref{eq:m0gluon2}) the local QCD sum rule~\eqref{eq:localsr} agrees with the LQCD results for $\lambda_p$ within uncertainties (cf.~Table~I of~\cite{Leinweber:1995fn}). The corresponding Borel window is $0.7 \, {\rm GeV} \leq \overbar{M} \leq 1 \, {\rm GeV}$. The central values and uncertainties of the parameters that enter the definitions of the twist-2 and twist-3 pion DAs can be found in~(\ref{eq:a2a4}),~(\ref{eq:f3piw3}) and~(\ref{eq:eta3}).

\begin{figure}[t]
\centering
\includegraphics[width=\textwidth]{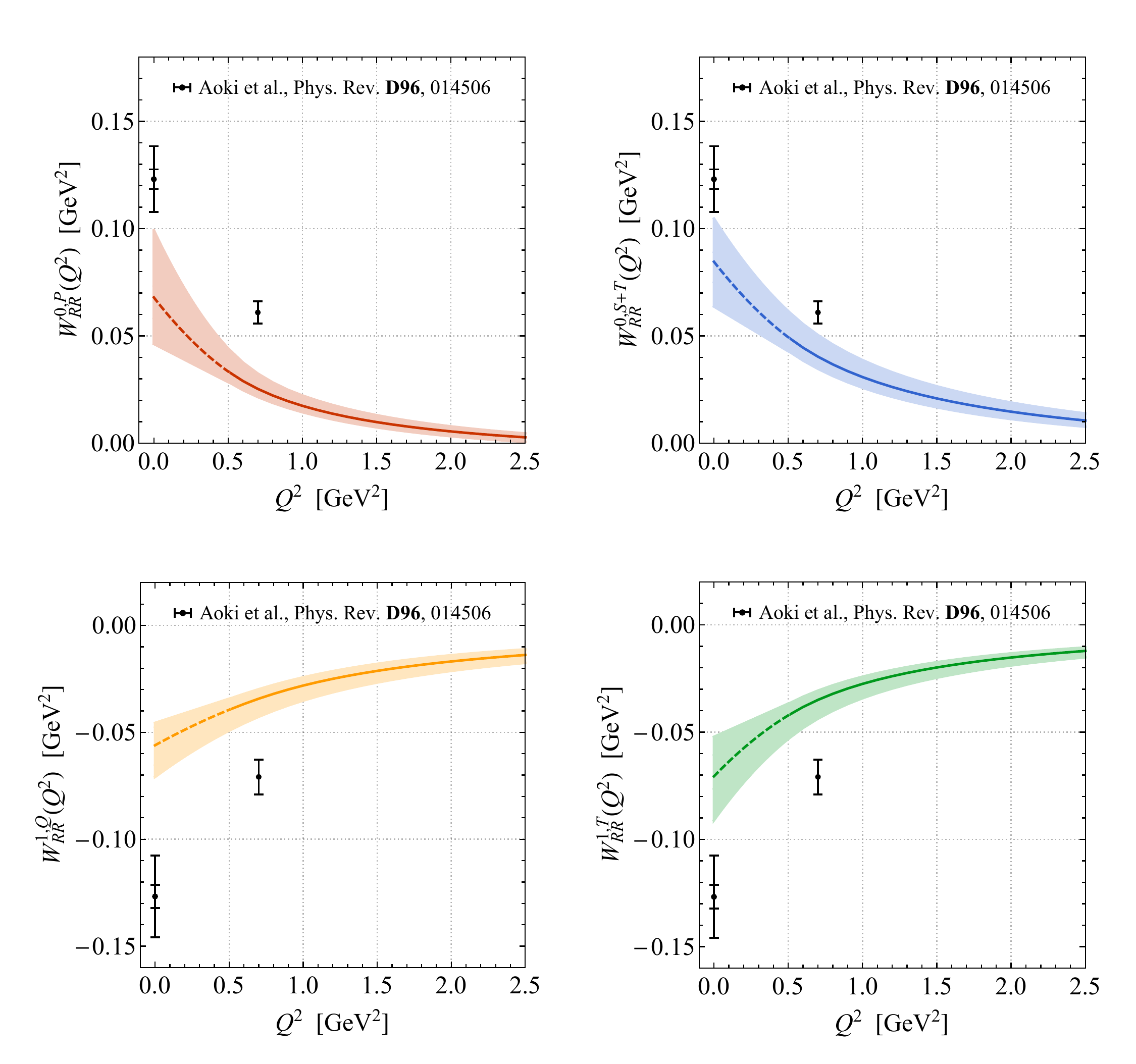}
\vspace{-5mm}
\caption{Form factors  $W^{n, \alpha}_{RR} (Q^2)$ as a function of $Q^2$. The coloured curves and bands correspond to the central values and uncertainties of the four independent LCSRs~(\ref{eq:ourLCSRs}). The predictions for $0.5 \, {\rm GeV}^2 \leq Q^2 \leq 2.5 \, {\rm GeV}^2$ are obtained by a direct calculation~(solid lines), while the predictions for $Q^2 \leq 0.5 \, {\rm GeV}^2$ are obtained by an extrapolation~(dashed lines). The black dots display the central values of the form factors calculated in LQCD~\cite{Aoki:2017puj}. The associated single (double) error bars represent statistical~(total) uncertainties. Consult the main text for further information.}
\label{fig:WRR}
\end{figure}

\begin{figure}[t]
\centering
\includegraphics[width=\textwidth]{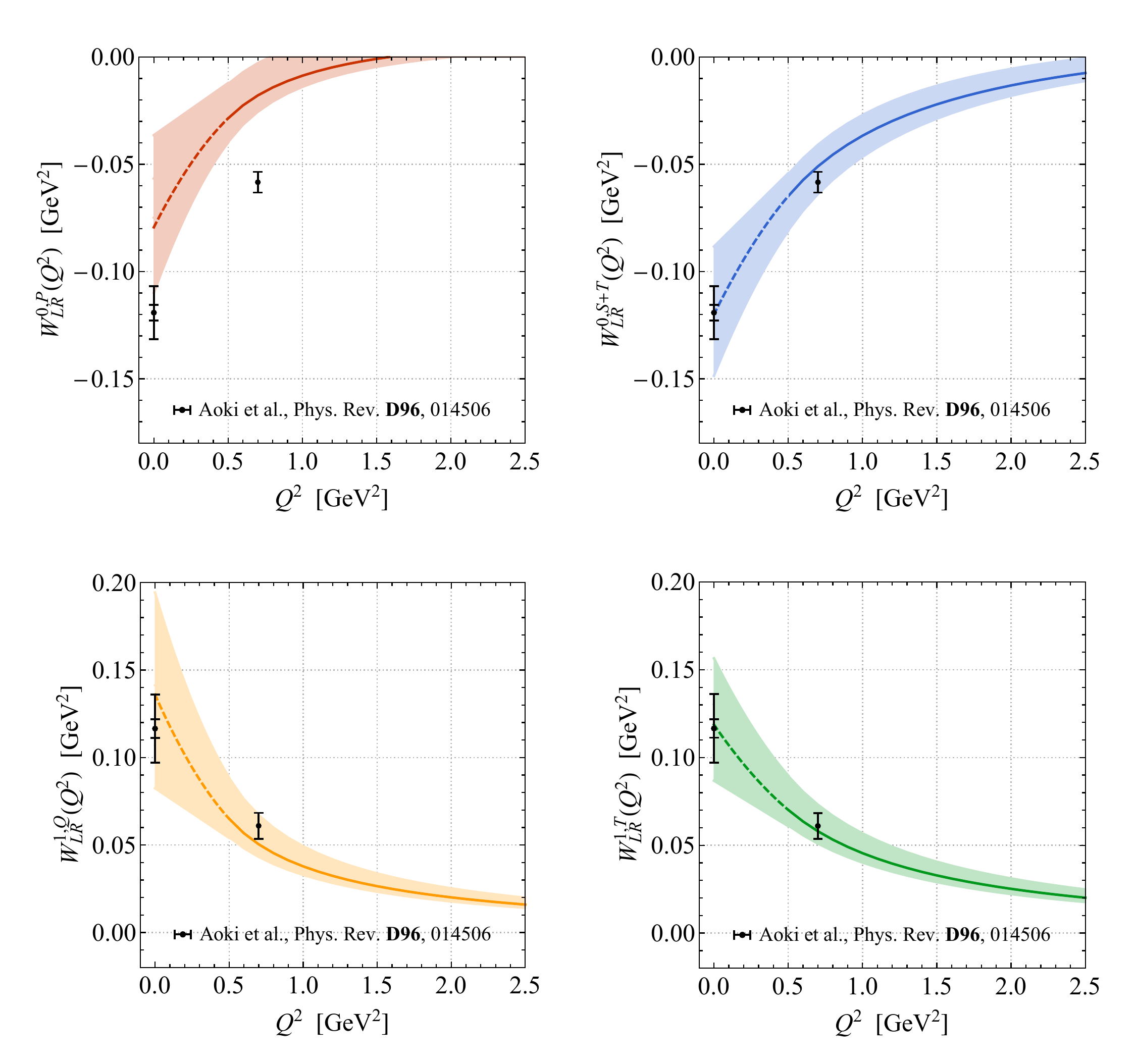}
\vspace{-5mm}
\caption{As Figure~\ref{fig:WRR} but for the form factors  $W^{n, \alpha}_{LR} (Q^2)$. }
\label{fig:WLR}
\end{figure}

After having explained how we choose the continuum thresholds  and the Borel mass windows and having specified the numerical values of the input parameters, we are in a position to present the results of our  LCSR analysis. Our results for the form factors~$W^{n, \alpha}_{RR} (Q^2)$ and $W^{n, \alpha}_{LR} (Q^2)$ are shown in Figure~\ref{fig:WRR} and Figure~\ref{fig:WLR} as coloured lines and bands, respectively. The predictions in the range $0.5 \, {\rm GeV}^2 \leq Q^2 \leq 2.5 \, {\rm GeV}^2$ result from a direct evaluation of~(\ref{eq:ourLCSRs}). The~solid curves correspond to the results obtained for the central values of the unphysical and physical parameters, while the bands reflects the corresponding theoretical uncertainties. The theoretical uncertainties are determined by varying all input parameters independently within their allowed ranges and adding individual uncertainties in quadrature.\footnote{Since we use the pion decay constant $f_\pi$ and the condensate $\braket{\bar{q}q}$ as input parameters the uncertainties of $\mu_\pi$ and $\rho_\pi$ $\big($cf.~(\ref{eq:mupirhopi})$\big)$ are not separately included when calculating the total uncertainties.}   For $Q^2 \leq 0.5 \, {\rm GeV}^2$ we instead rely on an extrapolation. Specifically, we consider both a linear and a quadratic fit in $Q^2$ to the LCSR form factors  $W^{n, \alpha}_{\Gamma \Gamma^\prime} (Q^2)$ evaluated in the vicinity of $Q^2 = 0.6 \, {\rm GeV}^2$, and take the smallest and largest values of the fits at each $Q^2$ to obtain the displayed uncertainty bands. The results of the quadratic fit to our central form-factor predictions are indicated as dashed lines. For comparison we also show the values of~$W^{n}_{\Gamma \Gamma^\prime} (Q^2)$ determined in the recent LQCD study~\cite{Aoki:2017puj}.  The numbers given in this article correspond to the $\overline{\rm MS}$ form factors evaluated at $2 \, {\rm GeV}$ and we use the two-loop~RG running~(cf.~\cite{Aoki:2006ib,Nihei:1994tx}) of~(\ref{eq:4foperators}) to evolve the form factors down to~$1 \, {\rm GeV}$. The shown single~(double) error bars represent the statistical~(total) uncertainties of the LQCD predictions.  Notice that the LQCD uncertainties  at the physical point,~i.e.~$Q^2 \simeq 0$, are dominantly of systematic origin.

As explained before, based on our study of the Borel windows we expect the LCSR prediction for~$W^{0,P}_{\Gamma \Gamma^\prime} (Q^2)$ to be less reliable than the other results because of the large contributions of heavy states. Indeed, comparing the results of $W^{0,P}_{\Gamma \Gamma^\prime} (Q^2)$ and $W^{0,S+T}_{\Gamma \Gamma^\prime} (Q^2)$ as shown in Figures~\ref{fig:WRR} and~\ref{fig:WLR}, one finds that $W^{0,S+T}_{\Gamma \Gamma^\prime} (Q^2)$ is closer to the LQCD predictions than $W^{0,P}_{\Gamma \Gamma^\prime} (Q^2)$ for both chirality combinations, and that  $W^{0,S+T}_{LR} (Q^2)$ itself agrees well with the LQCD calculation within uncertainties. One also observes from~Figure~\ref{fig:WRR}  that the LCSR predictions for the modulus of $W^{n,\alpha}_{RR} (Q^2)$ tend to undershoot the LQCD results. An~exhaustive comparison to the shown LQCD results for $Q^2 \gtrsim 0.5 \,{\rm GeV}^2$ would  require knowledge about the systematic uncertainties of the LQCD calculations for non-zero $Q^2$. A full error budget is  in Tables 4 and 5 of the work~\cite{Aoki:2017puj} however provided only  for~$Q^2 \simeq 0$.  Notice that  if the systematic  uncertainties at $Q^2 \gtrsim 0.5 \,{\rm GeV}^2$ were comparable to the systematic uncertainties at $Q^2 \simeq 0$, our LCSR results might in fact overlap with the displayed~LQCD~predictions for $Q^2 \gtrsim 0.5 \,{\rm GeV}^2$.  

The observed differences between the LCSR and the LQCD results may be related to higher-twist effects. In order to examine this issue, we have calculated the twist-4 corrections to~\eqref{eq:DA2}, which is the only two-particle twist-4 correction~\cite{Braun:1989iv},  using the hadronic input parameters provided in~\cite{Ball:2006wn}. We find that for $Q^2= 0.5 \, {\rm GeV}^2$ the relative corrections to the values of the form factors shown in Figure~\ref{fig:WRR} amount to $38\%$ for~$W^{0,P}_{RR}$, $5\%$ for~$W^{0,S+T}_{RR}$, $51\%$ for~$W^{1,Q}_{RR}$ and $26\%$ for~$W^{1,T}_{RR}$. Other twist-4 corrections to the LCSRs arise from additional three-particle~DAs~(see~\cite{Ball:1998je,Ball:2004ye} for details), and depending on their size and sign the actual effect of twist-4 corrections may be notably different from the numbers quoted here. Nevertheless, the corrections we have computed are larger for the vectorial structures than for the scalar and tensor structure. This could explain why our LCSR calculation of~$W^{n,\alpha}_{RR}(Q^2)$ seems to work better for  $\alpha=S+T,T$ than for~$\alpha=P,Q$.  As a comparison, the two-particle twist-4 contributions to the form factor values shown in Figure~\ref{fig:WLR} amount to $22\%$ for~$W^{0,P}_{LR}$, $7\%$ for~$W^{0,S+T}_{LR}$, $12\%$ for~$W^{1,Q}_{LR}$ and $31\%$ for~$W^{1,T}_{LR}$  at $Q^2= 0.5 \, {\rm GeV}^2$.  In this case the tensor structure receives a larger correction than the vectorial structures, but overall the twist-4 corrections seem to be better under control for~$\Gamma \Gamma^\prime =LR$ than for~$\Gamma \Gamma^\prime =RR$. This may  explain why the LCSR predictions for~$W^{n,\alpha}_{LR}(Q^2)$ are in general in good agreement with the LQCD results. In~conclusion, we expect that uncertainties due to higher twist are minor  for~$Q^2 \gtrsim 1 \, {\rm GeV}^2$, while in the range $0.5 \, {\rm GeV}^2 \lesssim Q^2 \lesssim 1 \, {\rm GeV}^2$, twist-4 corrections may in the case~$\Gamma \Gamma^\prime =RR$ account for the differences between our~LCSR predictions and the corresponding LQCD results. Notice that on general grounds one would expect that~the total uncertainties of the LCSRs become  larger for decreasing values of~$Q^2$, because the power suppression in $\Lambda_{\rm QCD}^2/Q^2$ of the light-cone expansion~(\ref{eq:twistexp})  starts to becomes ineffective.  In the plots of Figures~\ref{fig:WRR} and~\ref{fig:WLR} this effect is mimicked by our extrapolation procedure that leads to larger total uncertainties for $Q^2 \lesssim 0.5 \, {\rm GeV}^2$.

\begin{figure}[t]
\centering
\includegraphics[width=0.975\textwidth]{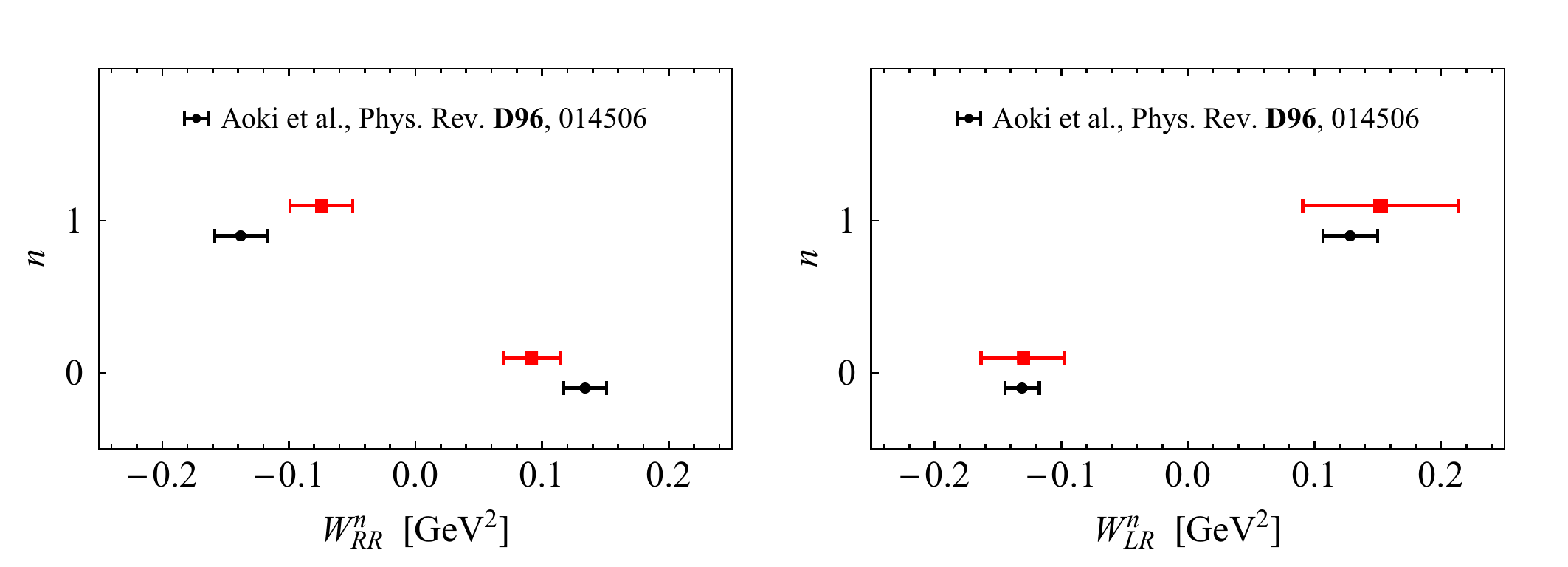}
\vspace{-1mm}
\caption{Comparison between the physical form factors $W^n_{RR}$ (left) and  $W^n_{LR}$ (right)  obtained by our LCSRs (red squares and error bars) and the state-of-the-art LQCD calculation (black~dots and error bars)~\cite{Aoki:2017puj}. The shown results correspond to the $\overline{\rm MS}$ scheme renormalised at $2 \, {\rm GeV}$. See~main text for additional details. }
\label{fig:wsummary}
\end{figure}

The physical form factors $W^0_{\Gamma \Gamma^\prime}  \equiv W^0_{\Gamma \Gamma^\prime} (Q^2 \simeq 0)$ can be extracted from both the~LCSR for~$W^{0,P}_{\Gamma \Gamma^\prime} (Q^2)$  and $W^{0,S+T}_{\Gamma \Gamma^\prime} (Q^2)$, while in the case of $W^1_{\Gamma \Gamma^\prime}   \equiv W^1_{\Gamma \Gamma^\prime} (Q^2 \simeq 0)$ one can consider the two independent combinations $W^{1,Q}_{\Gamma \Gamma^\prime} (Q^2)$  and~$W^{1,T}_{\Gamma \Gamma^\prime} (Q^2)$. Since we believe  that the~LCSR for~$W^{0,P}_{\Gamma \Gamma^\prime} (Q^2)$ is unreliable, we determine $W^0_{\Gamma \Gamma^\prime}$ from the full range of solutions for~$W^{0,S+T}_{\Gamma \Gamma^\prime} (0)$.  The prediction for the form factor~$W^1_{\Gamma \Gamma^\prime}$ is  instead obtained  from the extrapolations leading to $W^{1,Q}_{\Gamma \Gamma^\prime} (0)$  and $W^{1,T}_{\Gamma \Gamma^\prime} (0)$, because   in this case the different LCSR  estimates result in quite similar numerical predictions (see~Figures~\ref{fig:WRR}~and~\ref{fig:WLR}). At a renormalisation scale of $1 \, {\rm GeV}$, we obtain in this way  the following central values and uncertainties:
\begin{align} 
W_{RR}^{0} & = \left ( 0.084 \pm 0.021 \right ) {\rm GeV}^2 \,, \hspace{-5mm} & W_{RR}^{1} & = \left ( -0.068 \pm 0.023 \right )  {\rm GeV}^2  \,, \hspace{2mm} \label{eq:WRRfinal1GeV} \\[2mm]
W_{LR}^{0} & = \left ( -0.118 \pm 0.030  \right ) {\rm GeV}^2 \,,   \hspace{-5mm}   & W_{LR}^{1} & = \left ( 0.14 \pm 0.06 \right )  {\rm GeV}^2   \label{eq:WLRfinal1GeV} \,.
\end{align}
Our~LCSR predictions have total uncertainties of around $(25-40)\%$. In Figure~\ref{fig:wsummary} we compare the results~(\ref{eq:WRRfinal1GeV})  and~(\ref{eq:WLRfinal1GeV}) evolved to $2 \, {\rm GeV}$ to the corresponding  LQCD predictions~\cite{Aoki:2017puj}. Notice that two-loop RG effects (see~\cite{Aoki:2006ib,Nihei:1994tx}) lead to an enhancement of the~LCSR results by 8.9\% and 9.9\%, respectively. From the two panels it is evident that while the LCSR approach does not achieve the  $(10-15)\%$ accuracy of the latest LQCD computations of the form factors $W^n_{\Gamma \Gamma^\prime}$, the overall agreement between our~LCSR predictions and the latest~LQCD results is quite compelling.

\section{Conclusions}
\label{sec:conclusions}

In our work, we have calculated the hadronic matrix elements of the full set of baryon-number violating dimension-six SMEFT operators~(\ref{eq:4foperators}) using LCSR techniques.  These hadronic matrix elements are needed to predict  the rates of the main proton decay modes in GUTs, where a proton decays into a pseudoscalar meson and an anti-lepton. Specifically, we have focused on the decay $p \to \pi^0 e^+$, and presented explicit  LCSR expressions  for the relevant form factors that include the leading contributions in the light-cone expansion, namely the twist-2 and twist-3 pion DAs (cf.~Appendix~\ref{app:formulaeLCSRs}).  We have performed a detailed study of the dependence of the LCSRs on both the unphysical (i.e.~the continuum threshold and the Borel~mass) and the physical (i.e.~the condensates and the pion DAs) parameters, and discussed the possible impact of twist-4 effects. This enabled us to provide results and estimate uncertainties for the form factors in the kinematical regime where the momentum transfer~$q$ from the proton to the pion is space-like,~i.e.~$Q^2 = -q^2 > 0$, and lies in the range $0.5 \, {\rm GeV}^2 \leq Q^2 \leq 2.5 \, {\rm GeV}^2$. We have then extrapolated our LCSR results to the physical point $Q^2 \simeq 0$ by means of both a linear and quadratic fit, including the spread of predictions in our uncertainty estimates. Our  analysis indicates that the~LCSR for  $W^{0,P}_{\Gamma \Gamma^\prime} (Q^2)$ is not reliable, and we therefore consider only $W^{0,S+T}_{\Gamma \Gamma^\prime} (Q^2)$, $W^{1,Q}_{\Gamma \Gamma^\prime} (Q^2)$ and $W^{1,T}_{\Gamma \Gamma^\prime} (Q^2)$ when determining the  final predictions for the physical form factors $W^n_{\Gamma \Gamma^\prime}$  with~$n = 0, 1$ and~$\Gamma \Gamma^\prime = RR, LR$ from the range of    different solutions  shown in~Figures~\ref{fig:WRR} and~\ref{fig:WLR}.

Our final results for~$W^n_{\Gamma \Gamma^\prime}$ can be found in~(\ref{eq:WRRfinal1GeV}) and~(\ref{eq:WLRfinal1GeV}), and the LCSR results are compared to the state-of-the-art LQCD predictions~\cite{Aoki:2017puj} in~Figure~\ref{fig:wsummary}. The uncertainties of the LCSR results amount to~$(25-40)\%$, while the total accuracy of the LQCD form factors is~$(10 - 15)\%$. In view of the inherent systematic uncertainties of LCSRs, it is not clear to which extent possible refinements of our calculations such as including higher-twist contributions or perturbative corrections would allow to increase the precision of~(\ref{eq:WRRfinal1GeV})~and~(\ref{eq:WLRfinal1GeV}). The~observed overall agreement between our and the latest LQCD form factors demonstrates however that~LCSRs can be successfully applied to the calculations of proton decay matrix elements, and that such computations can achieve a precision that is better than the methods that have been developed in the '80s to estimate proton decay rates. 

This gives us confidence  that with the help of the LCSR techniques developed in this article  it should be possible to obtain (at least) order-of-magnitude estimates for the  hadronic matrix elements that appear  in certain three-body proton decays. One possible application is the decay mode $p \rightarrow \pi^0\, e^+ + G$ with~$G$~denoting a  graviton. This channel is expected~to~be the dominant proton decay mode in  the effective theory of gravity coupled to the~SM aka~GRSMEFT~\cite{Ruhdorfer:2019qmk,Durieux:2019siw},  since the two-body transition $p \to e^+ + G$ is forbidden by angular momentum conservation. LQCD calculations of three-body proton decay processes at arbitrary kinematics seem to be in reach in the coming years (see~\cite{Cirigliano:2019jig}~for a discussion), but it remains to be seen which accuracy such computations can initially achieve. The calculation of hadronic matrix elements for three-body proton decay modes utilising LCSRs therefore seems to be a worthwhile undertaking, and our work provides the blueprints for such future studies. 

\acknowledgments AH would like to thank Javi Serra for useful discussions on the topic. The analytical calculations in this article were performed with the help of {\tt FeynCalc}~\cite{Mertig:1990an,Shtabovenko:2016sxi,Shtabovenko:2020gxv}. Some of the Dirac traces were cross-checked against {\tt Tracer}~\cite{Jamin:1991dp}. The Feynman diagrams were created with the {\tt LaTeX} package {\tt feynMF}~\cite{Ohl:1995kr}.

\appendix

\section{Pion DAs}
\label{app:DA}

We use the following expressions for the pion DAs including terms proportional to the pion mass, which have been derived in~\cite{Ball:1998je} (and~\cite{Braun:1989iv} in the chiral limit) with the help of a conformal expansion. One has
\begin{align}
& \phi^{(2)}(u,\mu) =  6 \hspace{0.25mm} u  \bar{u} \left[ 1+ a_2(\mu) \hspace{0.25mm} C_2^{(3/2)}(\zeta) + a_4(\mu) \hspace{0.25mm} C_4^{(3/2)} (\zeta) \right] \,, \\[3mm]
&  \phi^{(3)}_p(u,\mu) =  1 + \left( 30 \hspace{0.25mm} \eta_3(\mu) - \frac{5}{2} \hspace{0.25mm} \rho_\pi^2 \right) C_2^{(1/2)} (\zeta) \notag \\[-2mm] \\[-2mm]
& \hspace{1.95cm} + \left(-3 \hspace{0.25mm} \eta_3(\mu)\hspace{0.25mm}  \omega_3(\mu) - \frac{27}{20} \hspace{0.25mm} \rho_\pi^2 - \frac{81}{10} \hspace{0.25mm} \rho_\pi^2 \hspace{0.25mm} a_2(\mu) \right) C_4^{(1/2)}(\zeta) \,,  \notag \\[3mm]
 & \phi^{(3)}_\sigma(u,\mu) = 6 \hspace{0.25mm} u \bar{u} \left[1 + \left( 5 \hspace{0.25mm} \eta_3(\mu) - \frac{1}{2} \hspace{0.25mm} \eta_3(\mu) \hspace{0.25mm} \omega_3(\mu) - \frac{7}{20} \hspace{0.25mm} \rho_\pi^2 - \frac{3}{5}\hspace{0.25mm}  \rho_\pi^2 \hspace{0.25mm} a_2(\mu) \right) \right] C_2^{(3/2)} (\zeta) \,,  \\[3mm]
 & \mathcal{T}^{(3)}\left(\alpha_d,\alpha_u,\alpha_g,\mu\right) =  360 \hspace{0.25mm} \eta_3(\mu) \hspace{0.25mm} \alpha_d \hspace{0.25mm} \alpha_u \hspace{0.25mm} \alpha_g^2 \left[1 +\frac{1}{2}\hspace{0.25mm}  \omega_3(\mu) \left(7 \alpha_g - 3\right)\right] \,, 
\end{align}
where the expansion in terms of the Gegenbauer polynomials $C_n^{(m)} (\zeta)$ with $\zeta \equiv 2 \hspace{0.25mm} u-1$ is truncated after $n=4$. The hadronic parameters that enter the above definitions depend on the renormalisation scale $\mu$  which we set equal to~$1 \, {\rm GeV}$ in our numerical analysis.  

We adopt the numerical values of the two Gegenbauer moments presented in~\cite{Khodjamirian:2011ub},
\begin{equation} \label{eq:a2a4}
a_2(1 \, {\rm GeV}) = 0.17 \pm 0.08 \,, \qquad  a_4(1 \, {\rm GeV}) = 0.06 \pm 0.10 \,,
\end{equation} 
where the moments are obtained by fitting sum rules for the electromagnetic pion form factor to the experimental data of~\cite{Huber:2008id}. For the numerical values of the other parameters we rely on the sum rules estimates of~\cite{Ball:2006wn}:
\begin{equation} \label{eq:f3piw3}
f_{3 \pi} (1 \, {\rm GeV}) = (0.45 \pm 0.15) \cdot 10^{-2} \, {\rm GeV}^2 \,, \qquad \omega_3(1 \, {\rm GeV}) = -1.5 \pm 0.7 \,.
\end{equation}
Using the definition $f_{3 \pi} (\mu) \equiv   f_\pi \hspace{0.25mm} \mu_\pi \hspace{0.25mm}  \eta_3 (\mu)$ together with~(\ref{eq:mupirhopi}) we then find,
\begin{equation} \label{eq:eta3}
\eta_3(1 \, {\rm GeV}) = 0.017 \pm 0.006 \,,
\end{equation}
where  the individual uncertainties are added in quadrature.

\section{Compendium of analytic formulas}
\label{app:formulae}

In the following we present a number of useful analytic formulas that enter our computations.  As a first step in obtaining the~LCSRs~\eqref{eq:ourLCSRs}, we have to perform the Fourier transformation, which amounts to solving integrals of the type
\begin{equation}
\int \! d^4x \,  e^{iPx} \, \frac{x_\mu x_\nu \ldots}{\left(x^2\right)^n} \,,\qquad \int \! d^4x \,  e^{iP_g x} \, \frac{x_\mu x_\nu\ldots}{\left(x^2\right)^n} \,,
\end{equation}
where the relevant momenta $P \equiv q+\bar{u}p_\pi$ and $P_g \equiv q+\alpha p_\pi$ with  $\alpha \equiv \alpha_u+u \alpha_g$ arise from combining the exponential factor in~\eqref{eq:corrfunc} with those of~\eqref{eq:DA2} to~\eqref{eq:DA3p}. The momentum dependence can be rewritten in terms of  $Q^2 = -q^2$ and $P_p^2 = -p^2_p$ using
\begin{align}
P^2 & = \left( \bar{u}p_p + u q\right)^2 = -\bar{u} P_p^2 - u \left(Q^2 + \bar{u} m_\pi^2\right) \,, \\[2mm]
P_g^2 & = \left( \alpha p_p + \bar{\alpha} q\right)^2 = -\alpha P_p^2 - \bar{\alpha} \left(Q^2 + \alpha m_\pi^2\right) \,,
\end{align}
where we employ the notation $\bar{x} \equiv 1-x$ for any variable $x$ throughout this section. 

The UV divergent Fourier integrals are carried out in dimensional regularisation. In~this step the poles and the scheme-dependent constants can be dropped in the sum-rule  calculation as long as we perform a Borel transformation in the end. Only inverse powers of the momenta, arising from finite integrals, and logarithms from the divergent integrals contribute to our sum rules. Moreover, only the imaginary parts of the correlation functions $\Pi^{\text{QCD},\alpha}_{\Gamma \Gamma^\prime}(s+i\epsilon,Q^2)$ enter the dispersion integrals~\eqref{eq:ourLCSRs}, and as explained in the main text we subtract the heavy contributions for $s > s_0$. We then perform a  Borel transform, which is defined by
\begin{equation}
\mathcal{B}_{P^2_p}\left[F (P^2_p ) \right] \equiv \lim_{n\rightarrow \infty} \frac{\left(M^2\right)^n}{(n-1)!} \left(-\frac{d}{d M^2}\right)^n F (n M^2 )  \,,
\label{eq:boreltrafo}
\end{equation}
for some function $F (P^2_p)$. In the final step the integration over $s$ is performed.

All the steps described above can be translated into certain replacement rules. For the logarithmic terms we find
\begin{equation}
\int^1_0 \! du \, f(u) \left(P^2\right)^n \ln \left(\frac{-P^2}{\mu^2}\right) \rightarrow - n! \int_0^\Delta \! du \, f(u) \, (\bar{u} M^2)^n  \, e (\tilde s)  \,  {\tilde E}_{n+1} (\tilde s) \,,
\end{equation}
with $\mu$ the renormalisation scale and $f (u)$ some function which depends on the momentum fraction $u$. Here we have introduced
\begin{align}
& \hspace{0.5cm} \Delta  \equiv \frac{s_0+Q^2+m_\pi^2}{2 m_\pi^2} \left( 1- \sqrt{1- \frac{4 m_\pi^2 s_0}{(s_0+Q^2+m_\pi^2)^2}}\right) \,, \label{eq:def1} \\[2mm]
e(s) & \equiv   e^{-\frac{s}{M^2}} \,, \qquad \tilde E_n (s) \equiv E_n \left( \frac{s_0-s}{M^2} \right) \,, \qquad \tilde{s}  \equiv \frac{u}{\bar{u}} \hspace{-0.25mm} \left (Q^2+\bar{u}m_\pi^2 \right ) \,,  \label{eq:def2}
\end{align}
where the definition of $E_n (x)$ can be found in~(\ref{eq:Enx}). The upper limit of the integration over~$u$ satisfies $\Delta(Q^2=0)=1=\Delta(s_0\rightarrow \infty)$ and $\Delta \leq 1$, and it arises because the dispersion integral only has support if $s_0 \geq \tilde{s}$. 

For terms involving three-particle DAs one furthermore has
\begin{equation}
\begin{split}
\int^1_0 \! D\alpha \, f(u,\alpha_u, \alpha_g) \left(P_g^2\right)^n \ln \left(\frac{-P_g^2}{\mu^2}\right) \rightarrow & - n! \int_0^1 D\alpha\, \theta \left( \alpha - \Delta_g \right) f(u,\alpha_u, \alpha_g)  \\[2mm] 
& \hspace{1cm} \times  (\alpha M^2)^n \, e ( \tilde{s}_g ) \, \tilde E_{n+1} \left( \tilde{s}_g \right) \,,
\end{split}
\end{equation}
where $ f(u,\alpha_u, \alpha_g) $ now depends on $u$, $\alpha_u$ and $\alpha_g$, and we have eliminated $\alpha_d=1-\alpha_u-\alpha_g$. We have furthermore used the abbreviation 
\begin{equation}
D\alpha\equiv du \hspace{0.5mm} d\alpha_u \hspace{0.5mm} d\alpha_g \hspace{0.5mm} \theta (1-\alpha_u-\alpha_g) \,, \label{eq:def3}
\end{equation}
for the integration measure.  The integration boundaries are now modified by the Heaviside~step function $\theta (x)$,
\begin{equation} \label{eq:def4}
\Delta_g \equiv \frac{s_0+Q^2-m_\pi^2}{2 m_\pi^2} \left( \sqrt{1+\frac{4 m_\pi^2 Q^2}{(s_0+Q^2-m_\pi^2)^2}}- 1 \right) \,, \qquad  \tilde{s}_g \equiv  \frac{\bar{\alpha}}{\alpha}  \hspace{-0.25mm}  \left (Q^2+\alpha m_\pi^2 \right ) \,, 
\end{equation}
where $\Delta_g(Q^2=0)=0=\Delta(s_0\rightarrow \infty)$ and $\Delta_g \geq 0$. For the non-divergent contributions appearing in our LCSRs, we find
\begin{align}
&\begin{aligned}
\int^1_0 \! du \, f(u) \, \frac{1}{P^2}  \phantom{} \rightarrow - \frac{1}{M^2}\int_0^\Delta \! du \, \frac{f(u)}{\bar{u}} \, e(\tilde s)  \,,
\end{aligned} \\[3mm]
&\begin{aligned}
\int^1_0 \! du \, f(u) \, \frac{1}{P^4}  \phantom{} \rightarrow  \phantom{} \frac{1}{M^4}\int_0^\Delta \! du \, \frac{f(u)}{\bar{u}^2} \, e(\tilde s) +  \frac{f(\Delta)}{M^2 \left( Q^2+\bar{\Delta}^2 m_\pi^2 \right)} \, e(s_0) \,,
\end{aligned} \\[3mm]
&\begin{aligned}
\int^1_0 \! du \, f(u) \, \frac{1}{P^6} \rightarrow & -\frac{1}{2 M^6}\int_0^\Delta \!du \, \frac{f(u)}{\bar{u}^3}\, e ( \tilde{s}) - \frac{f(\Delta)}{2 M^4 \bar{\Delta} \left(Q^2+\bar{\Delta}^2 m_\pi^2\right)} \, e(s_0)  \\[2mm]
& - \frac{\bar{\Delta}^2}{2 M^2 \left(Q^2+\bar{\Delta}^2 m_\pi^2\right)} \,  e(s_0) \, \frac{\partial}{\partial \Delta} \frac{f(\Delta)}{\bar{\Delta} (Q^2+\bar{\Delta}^2 m_\pi^2)} \,,
\end{aligned}
\end{align}
and for the three-particle integrals:
\begin{align}
&\begin{aligned}
\int^1_0 \! D\alpha \, f(u,\alpha_u, \alpha_g) \, \frac{1}{P_g^2}  \phantom{} \rightarrow - \frac{1}{M^2} \int_0^1 \!  \frac{D\alpha}{\alpha} \, \theta \left( \alpha - \Delta_g \right)  f(u,\alpha_u, \alpha_g)   \, e (\tilde s_g)  \,,
\end{aligned} \\[3mm]
&\begin{aligned}
\int^1_0 \! D\alpha \, f(u,\alpha_u, \alpha_g) \, \frac{1}{P_g^4} \rightarrow &  \phantom{i} \frac{1}{M^4} \int_0^1  \! \frac{D\alpha}{\alpha^2} \, \theta \left( \alpha - \Delta_g \right)  f(u,\alpha_u, \alpha_g) \, e (\tilde s_g)  \\[2mm]
& + \frac{1}{M^2} \int_0^1 \! du \, d\alpha_g \, \theta \left( 1- \bar{u} \alpha_g - \Delta_g \right) e (s_0)  \ \\[2mm]
& \quad \times \frac{ f(u, \Delta_g - u \alpha_g , \alpha_g)}{Q^2+\Delta_g^2 m_\pi^2}  \,.
\end{aligned} 
\end{align}

\section{Analytic results for LCSRs}
\label{app:formulaeLCSRs}

In this appendix, we  provide  the analytic expressions for the QCD correlation functions that appear on the right-hand side of the LCSRs~\eqref{eq:ourLCSRs} --- the integrations over the momentum fractions  have to be calculated numerically. The hat on the functions $\hat{\Pi}^{\text{QCD}, \alpha}_{\Gamma \Gamma^\prime}$ indicates that we have subtracted the contributions of heavy states before taking the Borel transform of the QCD results. We obtain
\begin{flalign}
\hat{\Pi}^{\text{QCD}, S}_{RR} = \frac{i \hspace{0.125mm} f_\pi}{32 \sqrt{2}} \, \Bigg\{ & \frac{m_0^2 \braket{\bar{q}q}}{3} \bigg[ \int_0^\Delta \! du \, \frac{\phi^{(2)}(u)}{\bar{u}^3 M^2} \, e(\tilde s) \left( Q^2+\bar{u}^2 m_\pi^2-\bar{u}M^2 \right)  &\notag\\[-2mm] \label{eq:C1} \\[-2mm]
& \phantom{xxxxxx} + \frac{\phi^{(2)}(\Delta)}{\bar{\Delta}} \, e(s_0)  \bigg] - \frac{3 \mu_\pi M^4}{\pi^2} \int_0^\Delta \! du \, \bar{u} \, \phi^{(3)}_p(u) \, e(\tilde s) \hspace{0.25mm} \tilde E_2 ( \tilde{s}  ) \Bigg\} \,,& \notag
\end{flalign}
\begin{flalign}
\hat{\Pi}^{\text{QCD}, P}_{RR} =  \frac{i \hspace{0.125mm} f_\pi m_p}{12 \sqrt{2}} \, \Bigg\{ & \frac{M^2}{8 \pi^2} \int_0^\Delta \! du\, \phi^{(2)}(u) \, e (\tilde s) \, 
 \Big [ \left( Q^2+\bar{u}^2 m_\pi^2 \right) \tilde E_1 ( \tilde{s} ) - 13\bar{u}M^2 \tilde E_2 ( \tilde{s} )\Big ] &\notag\\[2mm]
& \hspace{-2.5cm} + \frac{\mu_\pi m_0^2 \braket{\bar{q}q}}{4 M^2}  \bigg[ \int_0^\Delta du \, \frac{\phi^{(3)}_p(u)}{\bar{u}} \, e(\tilde{s}) +   \frac{\bar{\Delta} M^2 \hspace{0.125mm} \phi^{(3)}_p(\Delta)}{Q^2 + \bar{\Delta}^2 m_\pi^2} \, e (s_0)  \bigg] &\notag\\[2mm]
&\hspace{-2.5cm} + \frac{\mu_\pi \braket{\bar{q}q}}{3 M^2} \left(1-\rho_\pi^2\right) \, \bigg \{ \int_0^\Delta \! du \, \frac{\phi^{(3)}_\sigma(u)}{\bar{u}^3}\, e( \tilde{s} )   &\notag\\[2mm]
& \hspace{-2.5cm} \phantom{xxxx} \times \bigg[ 2 \bar{u}^2 M^2 \left(1- \frac{m_0^2}{12\bar{u}M^2} \right) - \bar{u} \left(Q^2+\bar{u}^2 m_\pi^2 \right) \left(1-\frac{m_0^2}{6\bar{u}M^2} \right) \bigg]  &\notag\\[2mm]
&  \hspace{-2.5cm} \phantom{xx} - M^2 \phi^{(3)}_\sigma(\Delta) \, e (s_0) \, \bigg(1- \frac{m_0^2}{6\bar{\Delta}M^2} \bigg) + \frac{\bar{\Delta} m_0^2 M^2}{6 \left(Q^2 + \bar{\Delta}^2 m_\pi^2 \right)} \, e (s_0) \,  \frac{\partial}{\partial \Delta} \phi^{(3)}_\sigma(\Delta) \bigg\} &\\[2mm]
& \hspace{-2.5cm} - \frac{\mu_\pi \braket{\bar{q}q}}{M^2} \bigg[ \int_0^1 \! D\alpha \,  \theta \left( \alpha-\Delta_g \right) \,\frac{\mathcal{T}^{(3)}\left(1-\alpha_u-\alpha_g, \alpha_u,\alpha_g \right)}{\alpha^3}  \,  e(\tilde{s}_g)&\notag\\[2mm]
& \hspace{-2.5cm}  \phantom{xxxx} \times \big[ 2\bar{u} Q^2 - 2\bar{u} \alpha M^2+ (1-4u) \alpha^2 m_\pi^2  \big] &\notag\\[2mm]
& \hspace{-2.5cm}  \phantom{xx}+ M^2 e(s_0) \int_0^1 \! du \, d\alpha_g \, \theta \left(1-\bar{u}\alpha_g -\Delta_g \right) \frac{\mathcal{T}^{(3)}\left(1-\bar{u}\alpha_g -\Delta_g, \Delta_g -u \alpha_g,\alpha_g \right)}{\Delta_g \left( Q^2+\Delta_g^2 m_\pi^2 \right)} &\notag\\[2mm]
& \hspace{-2.5cm}  \phantom{xxxx} \times \left[ 2 \bar{u}Q^2 + (1-4u) \Delta_g^2 m_\pi^2 \right] \bigg ] \Bigg\} \,, \notag&
\end{flalign}
\begin{flalign}
\hat{\Pi}^{\text{QCD},Q}_{RR} = \frac{i \hspace{0.125mm}f_\pi m_p}{12 \sqrt{2}} \, \Bigg\{ & \frac{M^2}{8 \pi^2} \int_0^\Delta \! du \, \frac{\phi^{(2)}(u)}{\bar{u}} \, e( \tilde s )
 \left[ u  \hspace{0.25mm} ( Q^2+\bar{u}^2 m_\pi^2 )  \hspace{0.25mm} \tilde E_1 ( \tilde{s} ) +\bar{u}  \hspace{0.25mm} ( 14- 13u )M^2  \tilde E_2 ( \tilde{s}) \right] &\notag\\[2mm]
& \hspace{-2.5cm} + \frac{\mu_\pi m_0^2 \braket{\bar{q}q}}{4 M^2}  \bigg[ \int_0^\Delta \! du \, \frac{u\, \phi^{(3)}_p(u)}{\bar{u}^2} \, e (\tilde s) +  \frac{\Delta M^2 \hspace{0.125mm}  \phi^{(3)}_p(\Delta)}{Q^2 + \bar{\Delta}^2 m_\pi^2} \, e (s_0) \bigg] &\notag\\[2mm]
& \hspace{-2.5cm}  + \frac{\mu_\pi \braket{\bar{q}q}}{3 M^2} \left(1-\rho_\pi^2\right) \, \bigg\{  \int_0^\Delta \! du \, \frac{\phi^{(3)}_\sigma(u)}{\bar{u}^4} \, e ( \tilde{s} ) & \notag \\[2mm]
& \hspace{-2.5cm} \phantom{xxxx} \times  \bigg[  \zeta \bar{u}  M^2 \left(1- \frac{m_0^2}{6 M^2} \frac{1+u}{\zeta} \right)  - u \bar{u} \left(Q^2+\bar{u}^2 m_\pi^2 \right) \left(1-\frac{m_0^2}{6\bar{u}M^2} \right) \bigg]  &\notag \\[2mm]
& \hspace{-2.5cm} \phantom{xx} - \frac{\Delta M^2 \phi^{(3)}_\sigma(\Delta)}{\bar{\Delta}} \, e (s_0) \Bigg(1- \frac{m_0^2}{6\bar{\Delta}M^2} \Bigg) + \frac{\Delta m_0^2 M^2}{6 \left(Q^2 + \bar{\Delta}^2 m_\pi^2 \right)} \, e (s_0) \, \frac{\partial}{\partial \Delta} \phi^{(3)}_\sigma(\Delta) \bigg \} &\\[2mm]
& \hspace{-2.5cm}+ \frac{\mu_\pi \braket{\bar{q}q}}{M^2} \bigg[ \int_0^1 \! D\alpha \,  \theta \left( \alpha-\Delta_g \right) \, \frac{\mathcal{T}^{(3)}\left(1-\alpha_u-\alpha_g, \alpha_u,\alpha_g \right)}{\alpha^3} \, e ( \tilde{s}_g) &\notag\\[2mm]
& \hspace{-2.5cm} \phantom{xxxx}  \times \big[ 2\bar{u} Q^2 - 2\bar{u} \alpha M^2 +\alpha \left( 1 +2u + (1-4u)  \hspace{0.25mm} \alpha  \right) m_\pi^2  \big]  +&\notag\\[2mm]
& \hspace{-2.5cm} \phantom{xx} + M^2 e ( s_0 ) \int_0^1 \! du \, d\alpha_g \, \theta \left(1-\bar{u}\alpha_g -\Delta_g \right) \, \frac{\mathcal{T}^{(3)}\left(1-\bar{u}\alpha_g -\Delta_g, \Delta_g -u \alpha_g,\alpha_g \right)}{\Delta_g \left( Q^2+\Delta_g^2 m_\pi^2 \right)} &\notag\\[2mm]
& \hspace{-2.5cm} \phantom{xxxx}  \times \left[ 2 \bar{u}Q^2 + \Delta_g \left( 1+2u+ (1-4u)  \Delta_g\right) m_\pi^2 \right] \bigg ] \Bigg\} \,, \notag&
\end{flalign}
\begin{flalign}
\hat{\Pi}^{\text{QCD}, T}_{RR} =  \frac{i \hspace{0.125mm} f_\pi m_p^2}{2 \sqrt{2}} \, \Bigg\{ & \frac{\braket{\bar{q}q}}{3} \bigg[ - \int_0^\Delta \! du \, \frac{\phi^{(2)}(u)}{\bar{u}}\, e(\tilde{s}) \left( 1-\frac{m_0^2}{6 \bar{u} M^2} \right) \notag \\[-2mm] \\[-2mm] 
& \hspace{-1.25cm}  + \frac{m_0^2  \hspace{0.25mm} \phi^{(2)}(\Delta)}{6 \left(Q^2+\bar{\Delta}^2 m_\pi^2 \right)} \, e(s_0) \bigg] + \frac{\mu_\pi M^2}{16\pi^2} \left(1-\rho_\pi^2\right) \int_0^\Delta \! du \,  \phi^{(3)}_\sigma(u) \, e (\tilde{s}) \, \tilde E_1 ( \tilde{s} ) \Bigg\} \,, \notag &
\end{flalign}
\begin{flalign}
\hat{\Pi}^{\text{QCD}, S}_{LR} =  \frac{i \hspace{0.125mm}f_\pi}{32 \sqrt{2}} \Bigg\{ & \frac{m_0^2 \braket{\bar{q}q}}{3} \bigg[\int_0^\Delta \! du \, \frac{\phi^{(2)}(u)}{\bar{u}^3 M^2} \, e ( \tilde{s} ) \left( Q^2+\bar{u}^2 m_\pi^2-\bar{u}M^2 \right) + \frac{\phi^{(2)}(\Delta)}{\bar{\Delta}} \, e ( s_0 ) \bigg] &\notag\\[2mm]
&+ \frac{4\mu_\pi M^4}{\pi^2} \int_0^\Delta du \, \bar{u} \, \phi^{(3)}_p(u) \, e ( \tilde{s} )\hspace{0.125mm} \tilde E_2 (\tilde{s}) &\notag\\[-2mm] \\[-2mm]
& + \frac{\mu_\pi M^2}{2 \pi^2} \int_0^1 D\alpha \,  \theta \left( \alpha-\Delta_g \right) \mathcal{T}^{(3)}\left(1-\alpha_u-\alpha_g, \alpha_u,\alpha_g \right) e (\tilde{s}_g )  &\notag\\[2mm]
& \phantom{xx} \times \left[ \frac{\left(Q^2+\alpha^2 m_\pi^2 \right)^2}{\alpha^3 M^2} - 6 m_\pi^2 \hspace{0.25mm} \tilde E_1 (\tilde{s}_g) \right]  \Bigg\} \,,& \notag
\end{flalign}
\begin{flalign}
\hat{\Pi}^{\text{QCD}, P}_{LR} =  \frac{i \hspace{0.125mm} f_\pi m_p}{12 \sqrt{2}} \Bigg\{ & \frac{3 M^4}{4 \pi^2} \int_0^\Delta \! du \, \bar{u}\, \phi^{(2)}(u) \, e (\tilde{s} ) \, \tilde E_2 (\tilde{s} ) + &\notag\\[2mm]
& \hspace{-2.5cm} + \frac{\mu_\pi m_0^2 \braket{\bar{q}q}}{4 M^2} \,  \bigg[ \int_0^\Delta \! du \, \frac{\phi^{(3)}_p(u)}{\bar{u}} \, e ( \tilde{s} ) + \frac{\bar{\Delta}M^2 \hspace{0.125mm} \phi^{(3)}_p(\Delta)}{Q^2 + \bar{\Delta}^2 m_\pi^2} \, e (s_0 ) \bigg] +&\notag\\[2mm]
&  \hspace{-2.5cm}  - \frac{\mu_\pi \braket{\bar{q}q}}{3 M^2} \left(1-\rho_\pi^2\right)  \,  \bigg\{ \int_0^\Delta \! du \, \frac{\phi^{(3)}_\sigma(u)}{\bar{u}^3} \, e (\tilde{s} )   &\notag\\[2mm]
& \hspace{-2.5cm}  \phantom{xxxx}  \times \bigg [ 4 \bar{u}^2 M^2 \left(1- \frac{7m_0^2}{96\bar{u}M^2} \right)  - 2 \bar{u} \left(Q^2+\bar{u}^2 m_\pi^2 \right) \left(1-\frac{7 m_0^2}{48\bar{u}M^2} \right) \bigg ] &\notag\\[2mm]
& \hspace{-2.5cm}  \phantom{xx} - 2 M^2 \phi^{(3)}_\sigma(\Delta) \, e (s_0 ) \bigg(1- \frac{7 m_0^2}{48\bar{\Delta}M^2} \bigg) + \frac{7\bar{\Delta} m_0^2 M^2}{24 \left(Q^2 + \bar{\Delta}^2 m_\pi^2 \right)} \, e (s_0) \, \frac{\partial}{\partial \Delta} \phi^{(3)}_\sigma(\Delta) \bigg \} &  \\[2mm] 
& \hspace{-2.5cm}   + \frac{\mu_\pi \braket{\bar{q}q}}{M^2} \bigg[ \int_0^1 D\alpha \,  \theta \left( \alpha-\Delta_g \right) \frac{\mathcal{T}^{(3)}\left(1-\alpha_u-\alpha_g, \alpha_u,\alpha_g \right)}{\alpha^3} \, e ( \tilde{s}_g ) &\notag\\[2mm]
& \hspace{-2.5cm}  \phantom{xxxx}  \times \big[ (1-4u) \hspace{0.125mm} Q^2 - (1-4u) \hspace{0.125mm} \alpha M^2 +  (5-8u) \hspace{0.125mm} \alpha^2 \hspace{0.125mm} m_\pi^2 \big ]  &\notag\\[2mm]
& \hspace{-2.5cm}  \phantom{xx}  + M^2 e(s_0) \int_0^1 \! du \, d\alpha_g \, \theta \left(1-\bar{u}\alpha_g -\Delta_g \right)&\notag\\[2mm]
& \hspace{-2.5cm}  \phantom{xxxx} \times  \frac{\mathcal{T}^{(3)}\left(1-\bar{u}\alpha_g -\Delta_g, \Delta_g -u \alpha_g,\alpha_g \right)}{\Delta_g \left( Q^2+\Delta_g^2 m_\pi^2 \right)}   \left[ (1-4u) \hspace{0.125mm}Q^2 +  (5-8u) \hspace{0.125mm} \Delta_g^2 \hspace{0.125mm} m_\pi^2 \right] \bigg ] \Bigg\} \,,& \notag
\end{flalign}
\begin{flalign}
\hat{\Pi}^{\text{QCD}, Q}_{LR} =  \frac{i \hspace{0.125mm} f_\pi m_p}{12 \sqrt{2}} \,  \Bigg\{ & -\frac{3 M^4}{4 \pi^2} \int_0^\Delta \! du \, \bar{u}\, \phi^{(2)}(u) \, e ( \tilde{s}) \, \tilde E_2 ( \tilde{s}) &\notag\\[2mm]
&  \hspace{-3cm} + \frac{\mu_\pi m_0^2 \braket{\bar{q}q}}{4 M^2} \,  \bigg[ \int_0^\Delta \! du \, \frac{u\, \phi^{(3)}_p(u)}{\bar{u}^2} \, e ( \tilde{s} ) +  \frac{ \Delta  M^2 \hspace{0.125mm}  \phi^{(3)}_p(\Delta)}{Q^2 +  \bar{\Delta}^2  m_\pi^2} \, e (s_0)  \bigg] &\notag\\[2mm]
& \hspace{-3cm} - \frac{\mu_\pi \braket{\bar{q}q}}{3 M^2} \left(1-\rho_\pi^2\right)  \bigg\{ \int_0^\Delta \! du \, \frac{\phi^{(3)}_\sigma(u)}{\bar{u}^4} \, e ( \tilde{s} )  &\notag\\[2mm]
& \hspace{-3cm} \phantom{xxxx}  \times \bigg[  2 \zeta \bar{u} M^2 \left(1- \frac{7m_0^2}{48 M^2} \frac{1+u}{\zeta} \right)  -2 u \bar{u} \left(Q^2+\bar{u}^2 m_\pi^2 \right) \left(1-\frac{7 m_0^2}{48\bar{u}M^2} \right) \bigg] &\notag\\[2mm]
& \hspace{-3cm}  \phantom{xx}- \frac{2 \Delta M^2 \phi^{(3)}_\sigma(\Delta)}{\bar{\Delta}} \, e (s_0) \bigg(1- \frac{7 m_0^2}{48\bar{\Delta}M^2} \bigg) + \frac{7 \Delta m_0^2 M^2}{24 \left(Q^2 + \bar{\Delta}^2 m_\pi^2 \right)} \, e(s_0) \, \frac{\partial}{\partial \Delta} \phi^{(3)}_\sigma(\Delta) \bigg \} & \\[2mm] 
& \hspace{-3cm} - \frac{\mu_\pi \braket{\bar{q}q}}{M^2} \bigg[ \int_0^1 D\alpha \,  \theta \left( \alpha-\Delta_g \right) \frac{\mathcal{T}^{(3)}\left(1-\alpha_u-\alpha_g, \alpha_u,\alpha_g \right)}{\alpha^3} \, e ( \tilde{s}_g)  &\notag\\[2mm]
& \hspace{-3cm}  \phantom{xxxx} \times \big[  (1-4u) \hspace{0.125mm} Q^2 -(1-4u) \hspace{0.125mm} \alpha M^2 -   \left(4 \bar{u}-(5-8u) \hspace{0.125mm} \alpha \right) \alpha \hspace{0.125mm} m_\pi^2 \big] &\notag\\[2mm]
& \hspace{-3cm}  \phantom{xx}  + M^2 e(s_0) \,  \int_0^1 \! du \, d\alpha_g \, \theta \left(1-\bar{u}\alpha_g -\Delta_g \right) \frac{\mathcal{T}^{(3)}\left(1-\bar{u}\alpha_g -\Delta_g, \Delta_g -u \alpha_g,\alpha_g \right)}{\Delta_g \left( Q^2+\Delta_g^2 m_\pi^2 \right)}  &\notag\\[2mm]
& \hspace{-3cm}  \phantom{xxxx} \times \left[ (1-4u) \hspace{0.125mm} Q^2 - \left(4 \bar{u}- (5-8u)\hspace{0.125mm} \Delta_g\right) \Delta_g   \hspace{0.125mm}  m_\pi^2 \right] \bigg ] \Bigg\} \,, \notag&
\end{flalign}
\begin{flalign}
\hat{\Pi}^{\text{QCD}, T}_{LR} =  \frac{i \hspace{0.125mm} f_\pi m_p^2}{2 \sqrt{2}} \, \Bigg\{ & \frac{\braket{\bar{q}q}}{3} \bigg[2 \int_0^\Delta \! du \, \frac{\phi^{(2)}(u)}{\bar{u}} \, e ( \tilde{s})  \left( 1- \frac{7 m_0^2}{48\bar{u}M^2} \right)    - \frac{7 m_0^2 \hspace{0.25mm} \phi^{(2)}(\Delta)}{24 \left(Q^2+{\bar \Delta}^2 m_\pi^2 \right) \, e (s_0 ) } \bigg] &\notag\\[2mm]
&- \frac{\mu_\pi M^2}{12 \pi^2}\left(1-\rho_\pi^2\right) \int_0^\Delta \! du \, \phi^{(3)}_\sigma(u) \, e( \tilde{s} )\, \tilde E_1 ( \tilde{s} ) &\notag\\[-2mm] \label{eq:C8} \\[-2mm]
& - \frac{\mu_\pi}{16 \pi^2} \int_0^1 D\alpha \,  \theta \left( \alpha-\Delta_g \right)\frac{\mathcal{T}^{(3)}\left(1-\alpha_u-\alpha_g, \alpha_u,\alpha_g \right)}{\alpha^2} \, e ( \tilde{s}_g ) &\notag\\[2mm]
& \phantom{xx} \times \zeta \left(Q^2+\alpha^2 m_\pi^2 \right) \Bigg\} \,.& \notag
\end{flalign}
Recall that  $\zeta = 2u -1$ and notice that we have used the definitions~(\ref{eq:def1}),~(\ref{eq:def2}),~(\ref{eq:def3}) and~(\ref{eq:def4}) to write the  QCD correlation functions in a compact form. We have furthermore suppressed the renormalisation scale dependence of the pion DAs. The analytic expressions for the DAs are collected in Appendix~\ref{app:DA}. Notice that since we have  neglected quark-mass effects  in~(\ref{eq:quarkprop})~and~(\ref{eq:quarkproplocal}), it would be consistent to set to zero  all terms proportional to  $m_\pi^2$  in the formulas~(\ref{eq:C1}) to (\ref{eq:C8}). While these contributions are in fact numerically small, it turns out that they always improve the agreement between the LCSR form factors calculated here and the LQCD form factors computed in~\cite{Aoki:2017puj}. We therefore included the~$m_\pi^2$ terms in the expressions provided above.


%

\end{document}